\documentstyle[floats,prd,aps,epsf]{revtex}

\draft
\begin{document}

\newcommand{\beq}{\begin{equation}}
\newcommand{\eeq}{\end{equation}}
\newcommand{\beqn}{\begin{eqnarray}}
\newcommand{\eeqn}{\end{eqnarray}}
\newcommand{\pa}{\partial}
\newcommand{\vp}{\varphi}
\newcommand{\varep}{\varepsilon}
\def\bI{\hbox{$\,I\!\!\!\!-$}}
\def\agt{\mathrel{\raise.3ex\hbox{$>$}\mkern-14mu\lower0.6ex\hbox{$\sim$}}}
\def\alt{\mathrel{\raise.3ex\hbox{$<$}\mkern-14mu\lower0.6ex\hbox{$\sim$}}}



\begin{center}
{\large\bf{Numerical evolution of secular bar-mode instability 
induced by the gravitational radiation reaction in
rapidly rotating neutron stars}}
~~\\
~~\\
Masaru Shibata $^1$ and Sigeyuki Karino $^2$
~~\\
~~\\
{\em $^1$ Graduate School of Arts and Sciences, 
University of Tokyo, Komaba, Meguro, Tokyo 153-8902, Japan \\
$^2$ 
SISSA, International School for Advanced Studies, via Beirut 2/4,
34013 Trieste, Italy
}
\end{center}

\begin{abstract}
~~~\\
The evolution of a nonaxisymmetric bar-mode perturbation of rapidly
rotating stars due to a secular instability induced by
gravitational wave emission is studied in post-Newtonian simulations  
taking into account gravitational radiation reaction. 
A polytropic equation of state with the polytropic 
index $n=1$ is adopted. The ratio of the rotational kinetic energy
to the gravitational potential energy $T/|W|$ 
is chosen in the range between 0.2 and 0.26. 
Numerical simulations were performed until the perturbation
grows to the nonlinear regime, and illustrate 
that the outcome after the secular instability
sets in is an ellipsoidal star of a moderately large ellipticity
$\agt 0.7$. A rapidly rotating protoneutron star may form 
such an ellipsoid, which is a candidate for strong emitter 
of gravitational waves for ground-based laser interferometric detectors. 
A possibility that effects of magnetic fields 
neglected in this work may modify the growth of the secular instability
is also mentioned. 
\end{abstract}
\pacs{04.25.Dm, 04.30.-w, 04.40.Dg}

\section{INTRODUCTION}

Rapidly rotating stars are subject to nonaxisymmetric 
rotational instabilities.  An exact treatment of these instabilities 
exists for incompressible and rigidly rotating equilibrium fluids in
Newtonian gravity \cite{CH69,TA78}. For these configurations, global 
rotational instabilities arise from nonradial toroidal modes 
$e^{im\varphi}$ ($m=1,2, \dots$) when $\beta\equiv T/|W|$ exceeds a 
certain critical value. Here $\varphi$ is the azimuthal coordinate and 
$T$ and $W$ are the rotational kinetic and gravitational potential 
energies (see Sec. II B for definition). There exist two different types of
mechanisms and corresponding time scales for 
the instabilities. One is the {\em secular} instability which,
with a given value of $m$, 
sets in for a value of $\beta$ larger than a critical value $\beta_s$
and can grow in the presence of some dissipative mechanism, like 
viscosity or gravitational radiation. The growth time is 
determined by the dissipative time scale, which is usually longer 
than the dynamical time scale of the system. By contrast, a {\em 
dynamical} instability, which sets in for a value of $\beta$
larger than a critical value $\beta_d (> \beta_s)$, is independent of any 
dissipative mechanisms, and the growth time is determined 
typically by the hydrodynamical time scale of the system. 
In this paper, we study the growth of a secularly unstable 
bar-mode for compressible stars whose growth
is induced by the gravitational radiation. 
Hereafter, we only focus on the $m=2$ bar-mode 
since it is the fastest growing mode in most of rapidly rotating stars. 
(With regard to counter-examples in a narrow parameter range,
see \cite{comins,LS} about secular instability.) 

The criterion for the onset of the secular bar-mode instability 
for compressible rotating stars has been extensively determined 
by linear perturbative analyses in Newtonian theory 
\cite{linear1,linear2,linear3,linear4,linear5,linear6,linear7,KE}, 
in post-Newtonian approximation \cite{linear8}, and 
in general relativity \cite{SF,YE}. 
These studies indicate that the value of $\beta_s$ for a bar-mode 
is $\approx 0.14$ in many cases, although it 
varies depending on the rotational law, equations of state, and 
effect of general relativity. In particular, the value of $\beta_s$ 
could be decreased significantly below 0.14 for highly differentially 
rotating stars \cite{KE} and for a compact general relativistic 
star \cite{SF}. However, it is not increased much beyond 0.14 to our 
knowledge. Thus, in the presence of gravitational radiation reaction, 
the bar-mode perturbation can grow for rapidly rotating and 
secularly unstable stars with $\beta > 0.14$. 

The secular instability by gravitational wave emission may be relevant 
for rapidly and differentially rotating 
{\it protoneutron stars} formed soon after supernova collapse \cite{LS}. 
To study the growth of a secularly unstable 
bar-mode perturbation to a nonlinear perturbative regime,
a longterm numerical simulation is necessary. 
However, such study has not been performed until quite recently even in 
the Newtonian or post-Newtonian gravity. 
(But, see \cite{miller} for a study of the incompressible case
in which the basic equations reduce to ordinary differential equations, 
and \cite{LTV} for a simulation of $r$ mode instability.) 

There are several evolutionary paths which may lead to the formation 
of rapidly and differentially rotating neutron stars with $\beta > 0.14$. 
Assuming mass and angular momentum conservation, 
$\beta$ increases approximately as $R^{-1}$ during stellar collapse 
where $R$ denotes the radius of the collapsing core. 
In rotating supernova collapse, 
the core contracts from $R \sim 2000$ km to 
a few 10 km (e.g., \cite{BM,YS,ZM,RMR} for Newtonian simulations 
and \cite{HD,SS} for general relativistic simulations), 
and hence, $\beta$ may increase by two orders of magnitude. 
In reality, a large fraction of the mass is ejected in the supernova 
explosion, and hence, the relation between $\beta$ and $R$ 
will not be as simple as expected (e.g., \cite{HD}). 
Nevertheless, the value of $\beta$ in the inner region
that forms the protoneutron star should 
increase. The radius of a rotating progenitor star that forms
a protoneutron star of radius $\sim 10$ km will be 
$\gg 100$ km, and hence, it is reasonable to expect that 
the value of $\beta$ will increase at least 
by one order of magnitude. Thus, a stellar core with the effective value of
$\beta \agt 10^{-2}$, which can be easily reached in differentially rotating
cases \cite{HD}, may yield rapidly rotating neutron stars 
which may reach the onset of secular or dynamical instability. 
Similar arguments hold for accretion induced collapse 
of white dwarfs to neutron stars \cite{ID} 
and for the merger of binary white dwarfs to neutron stars. 

Only if the value of $\beta$ is in a range between $\beta_s$ and
$\beta_d$, a rapidly rotating neutron star is secularly unstable 
but dynamically stable. Thus, it may be more likely that 
a rapidly rotating protoneutron star is dynamically unstable.
The fate of dynamically unstable rotating stars against 
a bar-mode perturbation has been studied 
widely so far not only in Newtonian theory 
\cite{dyn1,dyn2,dyn3,dyn4,SKE} but also in general relativity \cite{dyn5}.
These studies have indicated that 
the value of $\beta_d$ is $\sim 0.27$ in many cases except for 
highly differentially rotating stars \cite{SKE} and that 
after the growth of the bar-mode perturbation, spiral arms are 
formed, and then, angular momentum is transported outward 
with mass-shedding. As a result, the angular momentum of 
around the center of the rotating star 
is decreased below the dynamically unstable limit
to be a slightly nonaxisymmetric star in which the value of
$\beta$ is smaller than $\beta_d$. However, $\beta$ remains still much
larger than $\beta_s$. Therefore, the relic of a dynamical instability 
can still be subject to a secular instability. 

In this paper, we report the results of numerical simulations 
for the secular bar-mode instability of rapidly and
differentially rotating protoneutron stars induced by
gravitational radiation.
The polytropic equation of state with the polytropic index $n=1$
is adopted to approximately model a protoneutron star. 
The hydrodynamic simulations are performed in 
a (0+2.5) post-Newtonian framework; i.e., Newtonian gravity and 
gravitational radiation reaction are taken into account \cite{BDS}. 
While protoneutron stars are fairly compact objects
with reasonably strong gravitational field, Newtonian theory is expected to 
describe them up to errors of the order $GM/Rc^2 \sim 0.1$--0.2 where
$G$, $c$, $M$, and $R$ are the gravitational constant, the 
speed of light, and mass and radius of the protoneutron stars, 
respectively. 

Although a longterm, stable, and accurate numerical simulations 
are now feasible (e.g., \cite{STU}), 
a simulation for the growth of a nonaxisymmetric secular instability 
in full general relativity is still formidable 
because of the following reasons.
(i) it is required to take a large computational
region that is extended to a wave zone to correctly compute gravitational
radiation reaction which is the key process in this issue, 
and (ii) it is required to perform a very longterm simulation because 
the growth time scale of a secularly unstable perturbation
is much longer (several orders of magnitude longer)
than the dynamical time scale of the system. These two facts imply 
that computational resources much larger than the present best ones 
are necessary to study this problem in fully general relativistic 
simulations. On the other hand, in a post-Newtonian simulation, 
(a) we can take into account the gravitational radiation reaction
simply adding a radiation reaction potential, and hence, we do not have to
take a large computational domain that is extended to a wave zone.
Furthermore, (b) the magnitude of radiation reaction force can be 
artificially increased to shorten the radiation reaction time scale as
short as a dynamical time scale. Specifically, we increase the magnitude of
the radiation reaction term by a factor $\sim 10$--100
in this paper to accelerate the
growth of a secularly unstable perturbation.
(In the (0+2.5) post-Newtonian framework, increasing the
radiation reaction force is equivalent to increasing the 
compactness of a star $GM/Rc^2$; see Sec. II A.) 
The facts (a) and (b) enable to reduce the computational time
significantly, and hence, the study of the secular instability for
compressible stars is feasible in a post-Newtonian framework
with current computational resources. 

The secular instability of rotating stars is also induced by
viscous effects in a different way from that by 
gravitational radiation \cite{CH69,DL,LS}: 
During the growth of the secular instability
induced by gravitational radiation, vorticity is conserved but 
angular momentum is dissipated. On the other hand, 
in the case of viscosity, the angular momentum is conserved but 
the vorticity is dissipated. If the viscous dissipation
time scale is as short as that of gravitational radiation,
two effects interfere each other and suppress the growth of the
secular instabilities \cite{DL}. An estimate suggests \cite{LS} that   
the viscous time scale in rapidly rotating protoneutron stars 
may not be as short as the dissipation time scale by gravitational
waves, although it is not very clear. In this paper,
we ignore the viscous effect as a first step and focus on the
secular instability induced by gravitational radiation. Simulation  
adding the viscous term (solving the Navier-Stokes equations) 
is an interesting subject remained in the future. 

The paper is organized as follows. In Sec. II, we describe
basic equations, initial conditions, and the methods in numerical analysis. 
In Sec. III, the numerical results are presented paying attention 
to the growth time of a secularly unstable perturbation, 
the outcome after the growth of the secular instability, and 
gravitational waves. Section IV is devoted to a summary. 
Throughout this paper, we use the geometrical units of $G=c=1$. 

\section{METHOD}

\subsection{Basic equations}

Numerical simulations are performed in a (0+2.5) post-Newtonian
framework \cite{BDS}. 
Here, ``0'' implies ``Newtonian'' order and ``2.5'' the 
lowest order radiation reaction effect due to mass-quadrupole 
gravitational wave emission. 
We adopt the following basic equations in Cartesian coordinates
\cite{BDS,ASF,RJ}: 
\beqn
&& {\pa \rho \over \pa t}+{\pa \rho v^j \over \pa x^j}=0,\\
&& {\pa \rho u_i \over \pa t}+{\pa (\rho u_i v^j +P \delta^{j}_i)
\over \pa x^j}=-\rho {\pa (\psi + \psi_R) \over \pa x^i},\\
&& {\pa \rho e \over \pa t}+{\pa (\rho e + P)v^i \over \pa x^i}=
-h_{ij} v^j {\pa P \over \pa x^i}
-\rho u_i {\pa (\psi + \psi_R) \over \pa x^i}.
\eeqn
The first, second, and third equations are the continuity, 
Euler, and energy equations, respectively. Here, $\rho$, $P$, $e$, $v^i$,
and $u_i$ are the baryon density, the 
pressure, the specific energy, the three velocity,
and the spatial components of
the four velocity. $e$ is the sum of the specific internal
energy $\varep$ and the specific kinetic energy as 
\beq
e \equiv \varep + {1 \over 2} u_i u_i. 
\eeq
$v^i$ and $u_i$ are related by
\beq
u_i=\tilde \gamma_{ij} v^j, 
\eeq
where $\tilde \gamma_{ij}=\delta_{ij}+h_{ij}$.
For consistency, we compute $v^j$ by
\beq
v^i=\tilde \gamma^{ij} u_j,
\eeq
where $\tilde \gamma^{ij}$ is the inverse of $\tilde \gamma_{ij}$. 

$\psi$, $\psi_R$, and $h_{ij}$ denote the Newtonian
potential, the radiation reaction potential, and the tracefree 
tensor component of the radiation reaction metric.
$\psi$ and $\psi_R$ are determined from the equations
\beqn
&& \Delta \psi=4\pi\rho,\\
&& \psi_R \equiv
{1 \over 2}\biggl(-\Phi+h_{ij}x^j{\pa \psi \over \pa x^i}\biggr),
\eeqn
where $\Phi$ obeys 
\beq
\Delta \Phi=4 \pi h_{ij} x^j {\pa \rho \over \pa x^i}. \label{eqphi}
\eeq
$h_{ij}$ can be related to the tracefree quadrupole moment $\bI_{ij}$ as
\beqn
h_{ij}=-{4G \over 5c^5}\epsilon {d^3 \bI_{ij} \over dt^3}, \label{hij}
\eeqn
where
\beqn
\bI_{ij} \equiv
\int d^3 x \rho \biggl(x^i x^j -{1 \over 3} r^2 \delta^{ij}\biggr). 
\eeqn
In Eq. (\ref{hij}), we recover $c$ to clarify that it appears
only in $h_{ij}$: In other terms of the basic equations, $c$ is not
included explicitly. $\epsilon$ is introduced to control the strength
of the radiation reaction force. 
Note that $\psi_R$ and $\Phi$ are also proportional to $\epsilon$.
Hence, the radiation reaction time scale will be proportional to
$\epsilon^{-1}$. 
To confirm that the growth time scale of the secular instability 
driven by gravitational radiation is indeed proportional to the inverse of
$\epsilon$, we performed simulations varying the value of $\epsilon$.

The order of magnitude for $d^3 \bI_{ij}/dt^3$ for rapidly rotating stars is
estimated as $\sim (GM/Rc^2)^{5/2}$. Thus, $h_{ij}$ is totally
proportional to $\epsilon (GM/Rc^2)^{5/2}$. This implies that 
increasing the value of $\epsilon$ is equivalent to increasing the
value of the compactness in the (0+2.5) post-Newtonian formalism.
We note that the compactness in this framework does not have
any meaning with regard to the strength of the self-gravity since
in this framework, $c$ appears only in the terms associated with the 
gravitational radiation reaction. 
The compactness here is only the indicator for the magnitude of
the radiation reaction force. 

The third time derivatives of $\bI_{ij}$ cannot be computed
accurately by the straightforward finite-differencing since 
we adopt a second-order accurate finite-differencing in time. 
For the accurate computation, we adopt the same method as 
that adopted in previous papers (e.g., \cite{SON,ONS}). In this
method, the time derivatives of the quadrupole moment are 
rewritten appropriately operating the time derivatives in the integral 
and using the equations of motion. Then, the third time
derivative of $\bI_{ij}$ is written 
\beqn
{d^3 \bI_{ij} \over dt^3}={\rm STF}\biggl[
2\int d^3x \biggl\{ 2P{\pa v^j \over \pa x^i}
-2\rho v^i {\pa \psi \over \pa x^j}+x^i {\pa \psi \over \pa x^j}
{\pa \rho v^k \over \pa x^k}-\rho x^i {\pa (\pa_t \psi) \over \pa x^j}
\biggr\}\biggr]+O(\epsilon),
\eeqn
where STF implies that the symmetric tracefree part should be extracted as
\beq
{\rm STF}[T_{ij}]={T_{ij} + T_{ji} \over 2}-{\delta_{ij} \over 3}T_{kk},
\eeq
and $T_{ij}$ is a tensor. 
The terms of order $\epsilon$ in $d^3\bI_{ij}/dt^3$ yield the terms of 
order $\epsilon^2$ in $h_{ij}$, and thus, it is neglected. 

During numerical simulation, we adopt the $\Gamma$-law equation of state 
\beqn
P=(\Gamma-1)\rho \varepsilon, 
\eeqn
where $\Gamma$ is the adiabatic index, and in this paper, we set 
$\Gamma=2$ to model a moderately stiff equation of state
for neutron stars as in \cite{LTV}.
With this equation of state, we can follow the formation of shocks 
that may form after the secular instability grows to a nonlinear regime. 

The hydrodynamic equations are solved using a so-called 
high-resolution shock-capturing scheme we have recently 
adopted in fully general relativistic simulations \cite{S03,STU}.
In this method, the transport terms in the hydrodynamic equations 
$\pa_i (\cdots)$ are computed by applying an approximate 
Riemann solver with third-order (piecewise parabolic) spatial interpolation. 
Detailed numerical methods with respect to the treatment of the
transport terms (in the framework of general relativity)
are described in \cite{S03}.
The time integration is done with the second-order Runge-Kutta method. 
Atmosphere of small density 
($\sim 10^{-6}$ of the central density $\rho_c$) outside stars is added
for using the shock-capturing scheme. Since the hydrodynamic equations
are written in the conservative form, the mass is conserved within
the digit of double precision. The conservation of 
energy and angular momentum is slightly violated due to numerical 
dissipation besides the loss by gravitational waves. 

In the (0+2.5) post-Newtonian formulation adopted here, 
we need to solve three Poisson equations for $\psi$, $\pa_t \psi$,
and $\Phi$. They are solved by a preconditioned conjugate 
gradient method, for which the details are given in \cite{ONS},
with boundary conditions
\beqn
&& \psi \rightarrow -{M \over r}-{3 \over 2 r^3} \bI_{ij} n^i n^j
+ O(r^{-4}),\\
&& \pa_t \psi \rightarrow -{3 \over 2r^3}{d \bI_{ij} \over dt} n^i n^j
+ O(r^{-4}),\\
&& \Phi \rightarrow -{3 \over 2r^3} C_{ij} n^i n^j + O(r^{-4}),
\eeqn
where $n^i$ is a radial unit vector $x^i/r$, 
$C_{ij}$ is a moment defined from 
the right-hand side of Eq. (\ref{eqphi}), and $M$ is the total
baryon mass defined by
\beq
M=\int d^3x \rho. 
\eeq

We adopt a fixed uniform grid with 
size $141\times 141\times 141$ in $x-y-z$, which covers a
region $-L \leq x \leq L$, $-L \leq y \leq L$, and $0 \leq z \leq L$
where $L$ is a constant. We assume a reflection symmetry
with respect to the equatorial plane, and 
set the grid spacing of $z$ to be half of that of $x$ and $y$. 
In all the simulations, the equatorial radius of rotating stars $R_e$
is set to be $5L/7$
(i.e., the equatorial radius is covered by 50 grid points). 
We also performed test simulations with size 
$113\times 113\times 113$ (i.e., the grid spacing becomes 1.25 times larger)
for several selected cases and confirmed that the results depend weakly 
on the grid resolution.
In the absence of radiation reaction, we checked the conservation of
the total energy and angular momentum. The magnitude of the error 
monotonically increases with the number of time steps and for a typical
simulation with $141 \times 141 \times 141$ grid size, 
after 30 central rotational periods ($P_0$), the violation of 
the total energy and angular momentum conservation 
is $\sim 5\%$ and $2\%$, respectively. We have checked that
these values are decreased at approximately second order with improving
the grid resolution. 
Even in the presence of gravitational radiation reaction,
the vorticity should be conserved. We have monitored its conservation 
and found that for a region of relatively high density 
$\agt 10^{-2}\rho_c$ where $\rho_c$ denotes the central density,
it is conserved within $\sim 10\%$ error. (Around the surface
of stars, the error is larger.) 
The violation of these conservation relations
is due to numerical dissipation or diffusion. 

Numerical simulations were performed on
FACOM VPP5000 in the data processing center of National Astronomical
Observatory of Japan.
For one numerical simulation with size $141 \times 141 \times 141$,
it takes about 8 CPU hours for 80,000 time steps using 48 processors.

\subsection{Initial conditions} 

A slightly perturbed rotating star is adopted 
as the initial condition for numerical simulation. Specifically, 
we first construct models of 
differentially rotating stars in axisymmetric equilibrium
with values of $\beta=T/|W|$ 
past the secular instability threshold \cite{KE}.
Here, in the notation of this paper, $T$ and $W$ are computed from 
\beqn
&&T \equiv {1 \over 2}\int d^3x \rho \varpi^2 \Omega^2,\\
&&W \equiv {1 \over 2}\int d^3x \rho \psi. 
\eeqn
The equilibrium state is computed using the polytropic equation of state,
\beq
P=K \rho^{\Gamma}, 
\eeq
where $K$ is the polytropic constant and $\Gamma=1+1/n$ with $n$ 
polytropic index. In this paper, we choose $\Gamma=2$ ($n=1$). 

Since a realistic velocity profile for protoneutron stars 
is poorly known (e.g., \cite{HD}), we give a plausible one. 
As the angular velocity profile $\Omega(\varpi)$
where $\varpi=\sqrt{x^2+y^2}$, 
we choose the so-called $j$-constant-like law as 
\beq
\Omega = {\Omega_0 A^2 \over \varpi^2 + A^2},\label{omegaj}
\eeq
where $A$ is a constant, and 
$\Omega_0$ the angular velocity at the rotational axis ($z$ axis). 
The parameter $A$ controls the steepness of the angular velocity profile:
For smaller values of $A$,
the profile is steeper and for $A \rightarrow \infty$, the 
rigid rotation is recovered. In the present work, the value of $A$ is 
chosen to be equal to the equatorial radius $R_e$. 
In this case, the ratio of $\Omega_0$ to $\Omega$ at the equatorial
surface is two, and hence, the star is moderately differentially rotating.
Numerical simulations have shown that such a rapidly rotating star
of moderate degree of differential rotation can be formed
from initial conditions with rapid and
moderately differential rotation \cite{HD}. 
As we have shown in previous papers \cite{SKE}, a star of a 
high degree of differential rotation, i.e., $\hat A \equiv  A/R_e \alt 0.5$ 
is dynamically unstable against nonaxisymmetric deformation 
even if it is not rapidly rotating with $\beta=O(0.01)$. 
However, with $\hat A =1$, the star is dynamically stable 
for $\beta \alt 0.265$ \cite{SKE,KE03}, 
and hence, the rapidly rotating stars with $0.2 \leq \beta \leq 0.26$ 
that we adopt in this paper (cf. Table I) 
can be subject only to the secular instability. 
We also note that only for $\hat A \alt 1$, a rotating 
star of a high value of $\beta \sim 0.25$ 
can be found for $\Gamma=2$ (e.g., \cite{KE}). Such a star 
which is secularly unstable but dynamically 
stable may be a plausible initial state of a rapidly rotating 
protoneutron star as mentioned in Sec. I. In addition, 
only for such a high value of $\beta$, 
the growth time scale of a secularly unstable 
mode is short enough to accurately follow 
the nonlinear growth of the perturbation. 

In terms of $\beta$ and $\hat A$, one rotating star is determined
for a given rotational profile and polytropic index. 
To specify a particular model, 
we could also specify the ratio of the polar radius $R_p$ to 
the equatorial radius $R_e$, i.e. $C_a=R_p/R_e$ instead of $\beta$.
For the equations 
of state and the angular velocity profiles that we adopt in this paper, 
the value of $C_a$ monotonically decreases with increasing $\beta$ 
for any given set of $\hat A$ and $n$. This is the reason that 
$C_a$ can be a substitute for $\beta$. 

\begin{figure}[tb]
\vspace*{-6mm}
\begin{center}
\epsfxsize=3.in
\leavevmode
(a)\epsffile{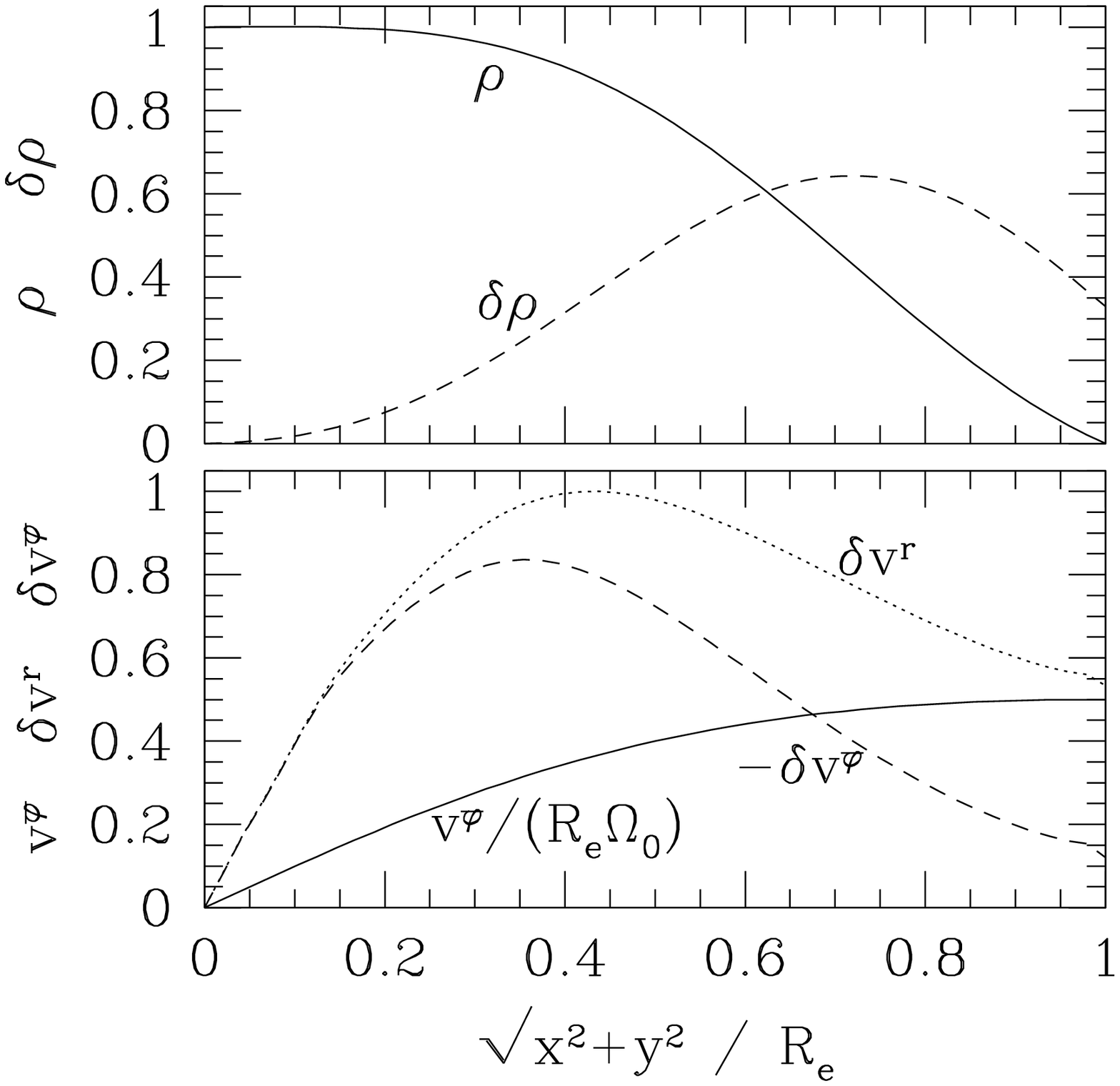}
\epsfxsize=3.5in
\leavevmode
~~(b)\epsffile{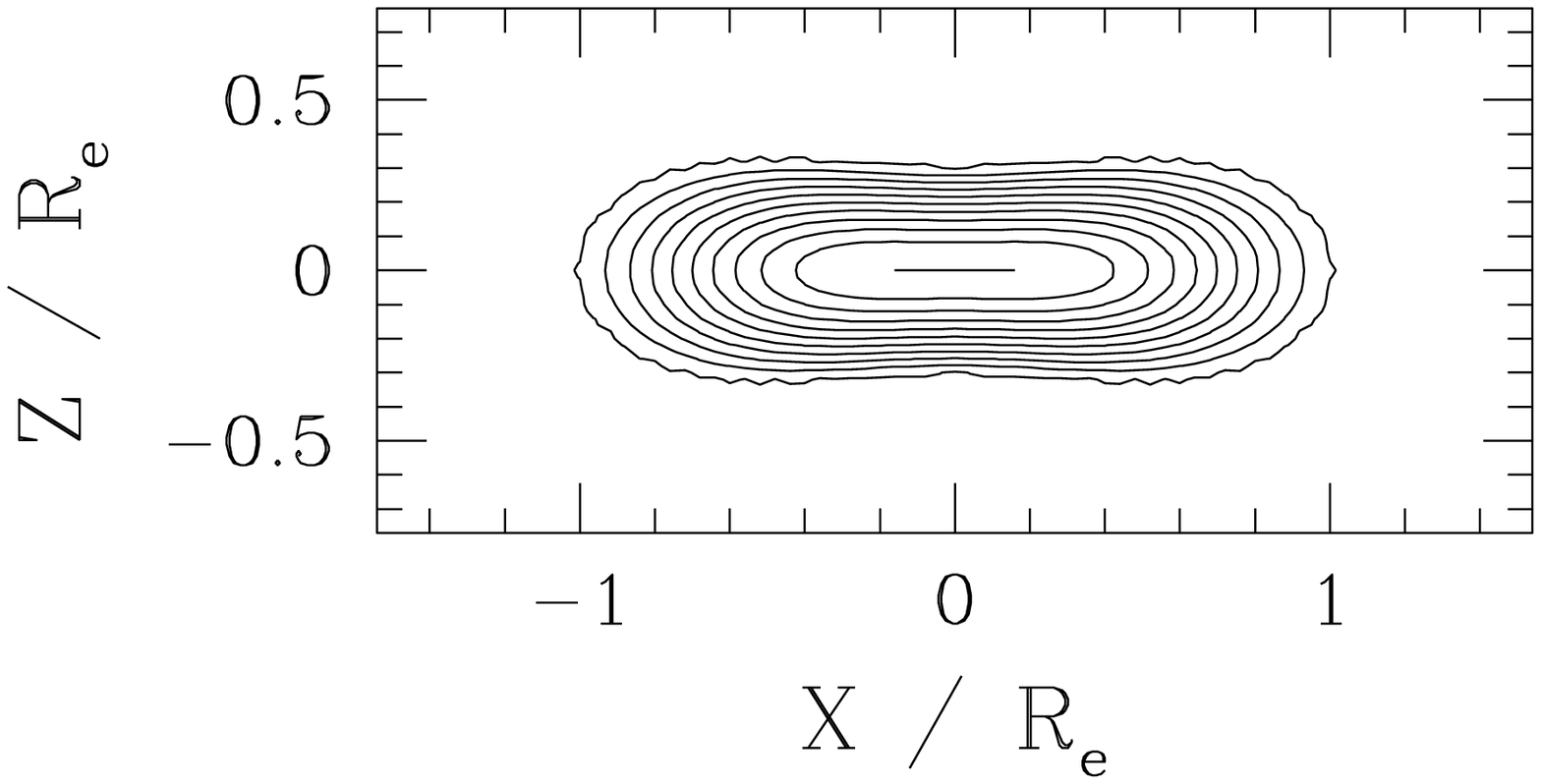}
\end{center}
\vspace*{-2mm}
\caption{(a) The profiles of the density, the rotational velocity,
and the perturbed quantities on the equatorial plane, and
(b) the density contour curves in $x$-$z$ plane for $C_a=0.3$.
The density and the rotational velocity are normalized
by $\rho_c$ and $\Omega_0 R_e$, respectively.
The amplitude of the perturbation is 
normalized so that the maximum value of $v^r$ is unity
in units of $G=\rho_c=R_e=1$. Here, $v^{\bar\varphi}=\varpi v^{\varphi}$. 
The density contour curves are drawn for $\rho/\rho_c=0.1 j$ 
for $j=0.01, 0.1, 1, 2, \cdots, 8, 9, 9.5$, and 10. 
\label{pert} }
\end{figure}

We initially add a secularly unstable perturbation
of a small amplitude on top of the equilibrium states.
The density and velocity profiles of an unstable mode adopted are those 
computed by Karino and Eriguchi \cite{KE} in a linear perturbation analysis. 
In their analysis, the perturbed quantities are expanded as 
\beqn
&&\delta \rho=\sum_m \delta \rho_m(r, \theta) \cos(m\varphi-\sigma t),\\
&&\delta v^r =\sum_m \delta v^r_m(r, \theta) \sin(m\varphi-\sigma t),\\
&&\delta v^{\bar \theta} =\sum_m \delta v^{\bar \theta}_m(r, \theta)
\sin(m\varphi-\sigma t),\\
&&\delta v^{\bar \varphi} =\sum_m
\delta v^{\bar \varphi}_m(r, \theta) \cos(m\varphi-\sigma t),
\eeqn
where $\sigma$ is the angular velocity of the perturbation. 
The eigenfunctions are these found by solving the perturbed equations 
for $\delta \rho_m$ and $\delta v^i_m$ 
in the two dimensional space of $r$ and $\theta$. Here, 
$v^{\bar\theta}$ and $v^{\bar\varphi}$ denote the components
associated with the orthonormal bases. 
We put an unstable mode of $m=2$ on a rotating equilibrium star 
at $t=0$. The profile of the perturbed quantities as well as
the density and the rotational velocity on the equatorial plane are
shown in Fig. \ref{pert} for $C_a=0.3$ for $m=2$, the only mode
considered here. Note that the profile of the perturbed quantities
is not as monotonic as that for rigidly rotating and incompressible
stars \cite{CH69}. Figure \ref{pert}(b) displays density contour
curves in $x$-$z$ plane for $C_a=0.3$ and shows that the 
rotating star is a flattened spheroid. 
In Table I, we also list several quantities of the perturbed rotating
stars adopted as 
the initial condition. Here, the energy and the angular momentum are
defined by
\beqn
&& E \equiv \int d^3 x \rho e + W,\\
&& J \equiv \int d^3 x \rho u_{\varphi}.
\eeqn

\begin{table}[tb]
\begin{center}
\begin{tabular}{|c|c|c|c|c|c|c|c|c|c|c|c|} \hline
$R_p/R_e$ & $T/|W|$ & $M/(\rho_c R_e^3)$ & $E/(M^2/R_e)$ &
$J/(M^{3/2}R_e^{1/2}) $ & $\Omega_0/\rho_c^{1/2}$ & $\sigma/\Omega_0$ &
$\sigma_+/\Omega_0$ & $I_{\varpi\varpi}/(M R_e^2)$ 
& $\hat \epsilon$ & $\tau/P_0$ & $\tau_{\rm anal}/P_0$  
\\ \hline\hline
0.30  & 0.256 & 0.567 & $-0.538$ & 0.366 & 1.34 & 0.569 & 0.914 &
0.288 & 0.25 & 3.5  & 5.9 \\ \hline
0.334 &  0.237 & 0.557 & $-0.546$ & 0.347 & 1.30 & 0.455 & 1.02 &
0.278 & 0.24 & 15 & 34 \\ \hline
0.35  & 0.229 & 0.565 & $-0.547$ & 0.339 & 1.29 & 0.419 & 1.06 &
0.275 & 0.25 & 31 & 60 \\ \hline
0.40  & 0.203 & 0.610 & $-0.547$ & 0.315 & 1.26 & 0.290 & 1.16 &
0.272 & 0.31 & --- & $510$ 
\\ \hline
\end{tabular}
\caption{Several quantities for secularly unstable rotating stars are 
shown in units of $G=R_e=\rho_c=1$ where $\rho_c$ is the central
density at $t=0$. $\sigma$ is the 
angular velocity of the unstable mode and $\tau$ is the
growth time scale of the secularly unstable mode 
found in numerical simulation for the
value of $\hat \epsilon \equiv \epsilon (GM/R_ec^2)^{5/2}$ given here 
in units of $P_0 \equiv 2\pi/\Omega_0$
(the rotational period along the $z$ axis).
Note that the value of $\hat \epsilon$ of a realistic neutron star
with the equatorial radius 10--20 km and the mass $1.4M_{\odot}$
is 0.018--0.003. ``---'' implies that we could not determine the value of
$\tau$ since it is too large. For all the models, 
the central density is normalized to be unity. 
}
\end{center}
\end{table}

The growth time of a secularly unstable mode is approximately proportional
to $\hat \epsilon^{-1}\equiv \epsilon^{-1}(GM/R_e c^2)^{-5/2}$. 
Thus, to accelerate the growth, we artificially increased the value of
$\hat \epsilon$ by a factor of 10--100. The values chosen in this work
are listed in Table I. In the following, we will refer to the
case listed in Table I as ``$\epsilon=1$'' case. 
To increase $\hat \epsilon$, 
we may either increase the value of $\epsilon$ or 
the value of compactness. As mentioned in Sec. II A, the compactness can be
rather arbitrarily increased in the (0+2.5) post-Newtonian framework.
In our choice of $\hat\epsilon$ for numerical computation, 
setting $\epsilon=1$ is equivalent to choosing very compact stars 
with $GM/R_e c^2 \sim 0.6$, 
which is about 3--6 times as large as that of a neutron star
or protoneutron star. We note that in general relativity, a star with 
$GM/R_e c^2 \sim 0.6$ collapses to a black hole. 
However, in the (0+2.5) post-Newtonian theory, 
a star of any value of $GM/R_ec^2$ is stable against collapse
for $\Gamma=2$. As a result, we can adopt such a large compactness 
of which the growth time scale of the secularly unstable mode is 
short enough to accurately follow its growth numerically. 
This is a great advantage in this post-Newtonian framework. 

\subsection{Method for analysis}

The growth of a nonaxisymmetric bar-mode perturbation is monitored 
using a distortion parameter defined as 
\beqn
\eta \equiv ( \eta_{+}^2 + \eta_{\times}^2)^{1/2}, 
\eeqn
where 
\beqn
&&\eta_+ \equiv {I_{xx} - I_{yy} \over I_{xx} + I_{yy}},\\
&&\eta_{\times} \equiv {2I_{xy} \over I_{xx} + I_{yy}},
\eeqn
and $I_{ij}~(i,j=x,y,z)$ denotes the quadrupole moment. 

Throughout this paper, we choose initial conditions 
in which $\eta \sim 0.01$ or 0.05 at $t=0$ 
by appropriately multiplying a constant 
to $\delta \rho_m$ and $\delta v^i_m$.
(Hereafter, we denote the initial value of $\eta$ as $\eta_0$.)
When the instability grows, $\eta$ should increase in proportional
to $e^{t/\tau}$ where $\tau$ denotes the growth time scale.
Thus, we fit the time evolution of $\eta$ and determine the
value of $\tau$. 

During numerical simulation, gravitational waves are computed in the
quadrupole formula \cite{MTW} in which ``$+$'' and ``$\times$''
polarization modes of the waveforms are defined as 
\beqn
h_+ \equiv {\ddot I_{xx} - \ddot I_{yy} \over r},~~~~
h_{\times} \equiv {2\ddot I_{xy} \over r},
\eeqn
and the luminosity and the angular momentum flux by 
\beqn
&&L_{\rm GW} \equiv {1 \over 5} \sum_{i,j} \bI_{ij}^{(3)}\bI_{ij}^{(3)}, \\
&&\dot J_{\rm GW} \equiv {2 \over 5} \sum_{j}
(\bI_{xj}^{(2)}\bI_{yj}^{(3)}-\bI_{yj}^{(2)}\bI_{xj}^{(3)}), 
\eeqn
where
\beqn
\ddot I_{ij}={d^2 I_{ij} \over dt^2},~~~~~
\bI_{ij}^{(n)}={d^n \bI_{ij} \over dt^n}~~(n=2, 3), 
\eeqn
and $r$ is the distance from the source to a detector. $h_+$ and $h_{\times}$
presented here are the waveforms observed along the rotational ($z$-) axis. 
Thus, effectively, they represent the maximum amplitude.

\section{NUMERICAL RESULTS}

\subsection{The growth of secularly unstable bar-mode perturbations}

\begin{figure}[tb]
\begin{center}
\epsfxsize=3.in
\leavevmode
(a) \epsffile{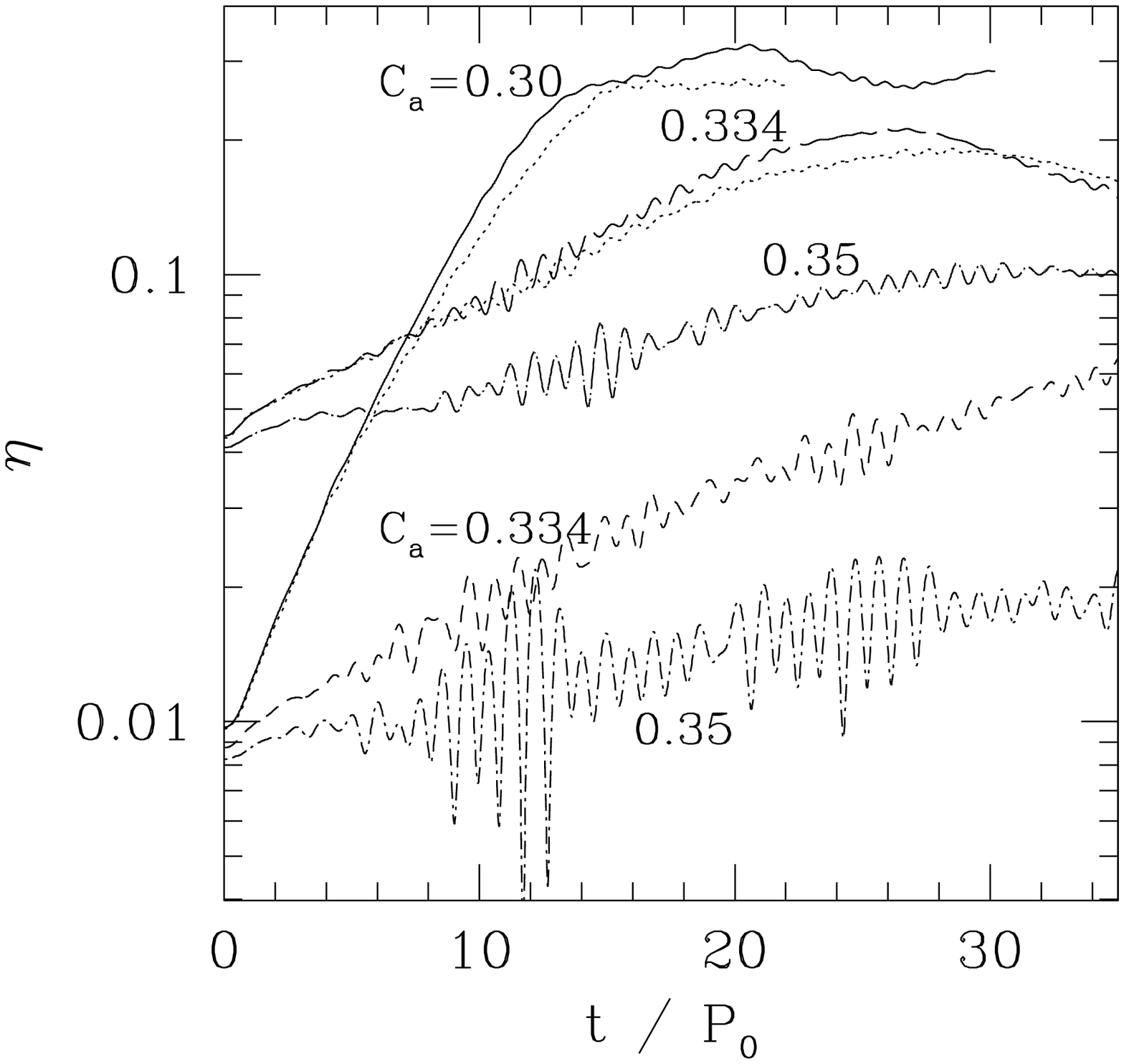}
\epsfxsize=3.in
\leavevmode
~~ (b)\epsffile{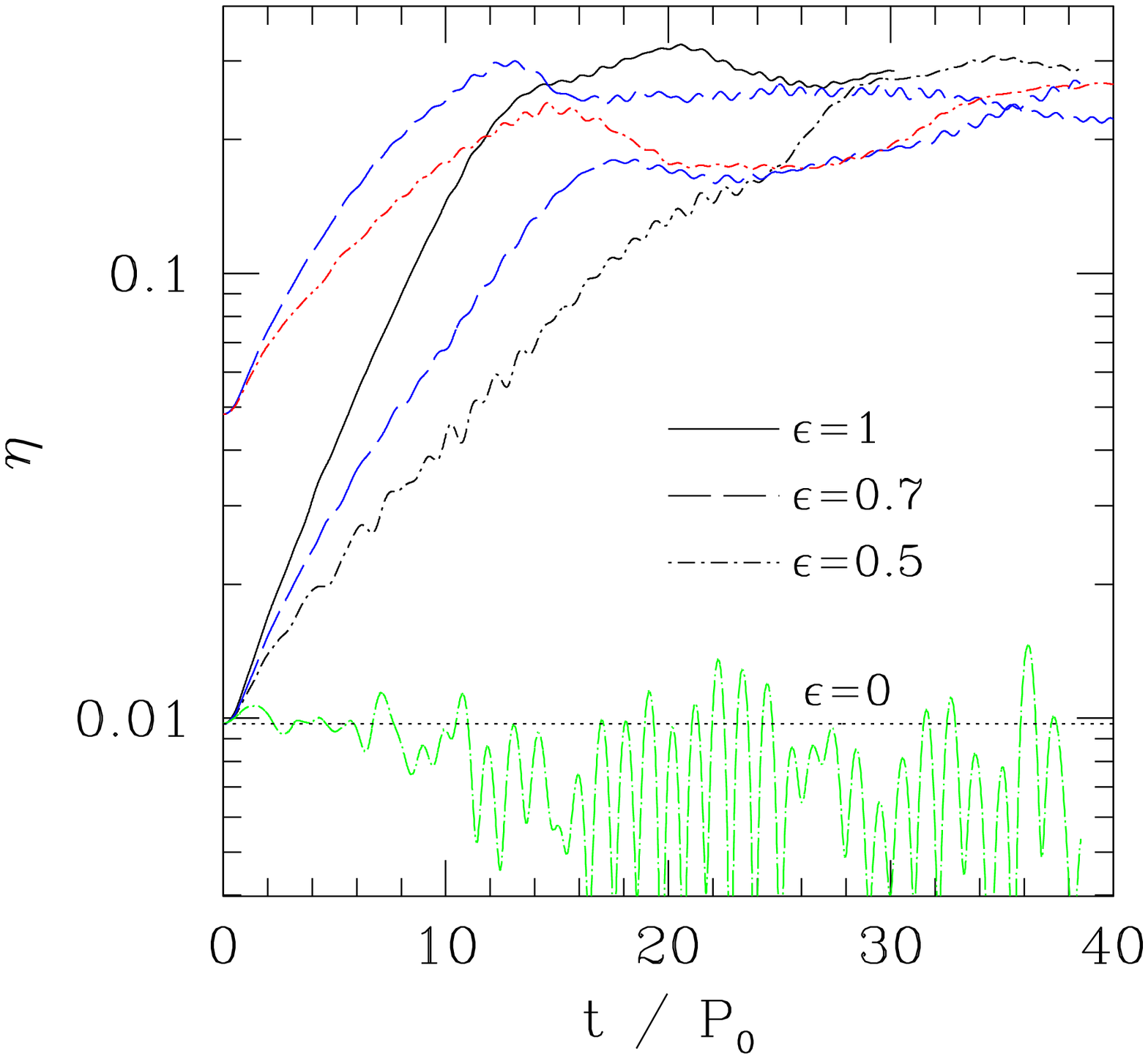}
\end{center}
\vspace*{-3mm}
\caption{(a) $\eta$ as a function of time for $C_a=0.3$ (solid curve),
0.334 (dashed curve), and 0.35 (dotted-dashed curve) with $\epsilon=1$
and $\eta_0 \sim 0.01$.
For comparison, the results with $\eta_0 \sim 0.04$ are
shown for $C_a=0.334$ (long-dashed curve) and
$C_a=0.35$ (dotted-long-dashed curve).
The dotted curves denote the results with a 
low-resolution grid for $C_a=0.30$ and $C_a=0.334$ with 
$\eta_0 \sim 0.04$. 
(b) $\eta$ as a function of time is shown for $C_a=0.3$ with
$\epsilon=1$ (solid curve), 0.7 (long-dashed curve),
0.5 (dotted-dashed curve), and 0 (no radiation reaction; dotted-long-dashed
curve). The dotted line denotes $\eta=\eta_0$.
For $\epsilon=0.5$ and 0.7, results with $\eta_0 \sim 0.05$ are
shown together. 
\label{FIG2} }
\end{figure}

Numerical simulations were performed for secularly unstable
stars with $C_a=0.3$, 0.334, 0.35, and 0.4.
For the case with $C_a \alt 0.52$ ($\beta \agt 0.14$), 
the star of $\hat A=1$ and $\Gamma=2$
is secularly unstable according to the result of a linear perturbation
analysis by Karino and Eriguchi \cite{KE}. In Fig. \ref{FIG2} (a),
we show $\eta$ as a function of time for $C_a=0.3, 0.334$, and 0.35
with $\epsilon=1$. 
For all cases, $\eta$ increases with time approximately as $e^{t/\tau}$.
The plausible reason for a slight deviation from the exponential form 
is that modes other than the secularly unstable ones are excited 
due to numerical noise or nonlinear coupling, which 
contribute to the growth and modulation of $\eta$
soon after we start the simulation. However,
they are not unstable modes, and hence, their amplitude does not
increase. As a result, after the amplitude of the unstable mode increases for
$t \agt 2 P_0$, the effect of other modes does not seem to be very large. 
From this reason, we determine the value of $\tau$ from the curve 
of $\eta$ for $t \agt 2 P_0$. More specifically, we use the data sets 
of $2 P_0 \leq t \leq 7 P_0$ for $C_a=0.3$, 
of $2 P_0 \leq t \leq 20 P_0$ for $C_a=0.334$ with $\eta_0 \sim 0.04$, and 
of $2 P_0 \leq t \leq 30 P_0$ for $C_a=0.35$ with $\eta_0 \sim 0.04$. 
We expect that the error for the evaluation of $\tau$ 
is $\sim 10$\% for $C_a=0.3$ and 0.334.
For $C_a=0.35$, the growth time scale is too long to follow the
late evolution with a sufficient accuracy. Moreover,
a modulation in $\eta$ prevents the accurate evaluation of
the growth time scale. Thus, the magnitude of the error in the
numerical result of $\tau$ for $C_a=0.35$ may be $> 10$\%. 

From the analysis, we determine $\tau/P_0 \approx 3.5$, 15, and
$31$ for $C_a=0.30$, 0.334, and 0.35 (see Table I). 
For $C_a=0.334$ and 0.35, the growth time is also evaluated
from the data with $\eta_0 \sim 0.01$, and we find that
the results agree approximately with those for $\eta_0 \sim 0.04$.
This shows that the growth time is independent of the value of $\eta_0$. 

According to a linear perturbative analysis for the bar-mode
secular instability of incompressible and rigidly rotating stars, 
the growth time of the secularly unstable perturbation
induced by emission of gravitational waves is \cite{linear1} 
\beqn
\tau_{\rm anal}={5 (\sigma_+ - \sigma) \over 2 I_{\varpi\varpi} \sigma^5},
\label{tau}
\eeqn
where $\sigma$ is the angular velocity of the secularly unstable mode,
$\sigma_+$ is that of its conjugate mode, which is a 
secularly stable one, and $I_{\varpi\varpi} \equiv I_{xx}+I_{yy}$.
Equation (\ref{tau}) is valid only for the incompressible 
and rigidly rotating case. For $n \not=0$, 
it may be an approximate formula \cite{LS}, but it is not very clear. 
Using the numerical results for $\sigma$ and $\sigma_+$
in a linear perturbative analysis by Karino and
Eriguchi \cite{KE}, the value of $\tau_{\rm anal}$ is evaluated and
listed in Table I. It is found that the value of
$\tau$ computed from the numerical results is systematically smaller than 
$\tau_{\rm anal}$: $\tau/\tau_{\rm anal}\sim 0.5$--0.6. 
The systematic disagreement will be due to the fact that Eq. (\ref{tau})
is valid only for the incompressible and rigidly rotating star. 
Indeed, the profile of the perturbed quantities is significantly
different from that for incompressible and rigidly rotating 
stars \cite{CH69}. On the other hand, the value of $\tau$ 
increases steeply as $\sigma/\Omega_0$ decreases by a 
similar manner to that described in Eq. (\ref{tau}). 
This indicates that the perturbation certainly grows due to 
the secular instability induced by gravitational radiation reaction. 

To check that $\tau$ is proportional to $\epsilon^{-1}$, 
in Fig. \ref{FIG2} (b), we show $\eta$ as a function of time for
$C_a=0.3$ with $\epsilon=1$ (solid curve), 0.7 (long-dashed curve),
0.5 (dotted-dashed curve), and 0 (no radiation reaction; dotted-long-dashed
curve). In these simulations, the compactness $GM/R_ec^2$ is fixed. 
For $\epsilon=1$, $\eta_0\sim 0.01$, but for 
$\epsilon=0.7$ and 0.5, $\eta_0$ is set to be $\sim 0.01$ and 0.05. 
For the case with no radiation reaction, the value of $\eta$
does not increase
but simply oscillates due to the growth of modes other than the
secularly unstable one probably by numerical noises and 
nonlinear coupling. This shows that the star with $C_a=0.3$
is dynamically stable.
(For $\hat A=1$ and $\Gamma=2$, only the stars with $C_a \alt 0.25$ are
dynamically unstable as demonstrated in \cite{SKE}.) 
The growth time $\tau$ is determined by the same method as that
adopted above. The analysis gives $\tau/P_0=3.5$, 4.9, and 7.4 
with an error of magnitude $\sim 10$\% for $\epsilon=1$, 0.7, and 0.5. 
Thus, $\tau/P_0$ is approximately proportional to $\epsilon^{-1}$.
We also note that $\tau$ also depends very weakly on 
the value of $\eta_0 \alt 0.05$. This implies that the amplitude of
the nonaxisymmetric perturbation initially given is small enough. 

Figure \ref{FIG2}(b) also indicates that the value of $\eta$ 
eventually reaches $\sim 0.3$ and remains between $\sim 0.2$ and $\sim 0.3$. 
This shows that an ellipsoid is formed. 
The maximum value of $\eta$ (hereafter $\eta_{\rm max}$) 
at the formation of the ellipsoid depends weakly on the magnitude of 
the parameter $\epsilon$. However, the shape of the curve
of $\eta(t)$ for $\eta \agt 0.1$ is fairly different. For $\epsilon=1$, 
$\eta$ increases almost monotonically to the maximum value, but
for $\epsilon=0.7$ and 0.5 with $\eta_0 \sim 0.01$,
the growth rate is varied at a time that 
$\eta$ becomes $\sim 0.1$. A plausible explanation for this difference
is that the radiation reaction time scale is too short to obtain a
qualitatively independent result of the value of $\epsilon$:
We recall that the radiation reaction force adopted here has an
invariant meaning only when the system is adiabatic, i.e., the system is
nearly periodic and the reaction time scale
is much longer than the period of a nonaxisymmetric oscillation.
For $C_a = 0.3$ with the compactness $\sim 0.6$, the reaction time scale 
is only a few times as long as the period of the 
nonaxisymmetric oscillation $\sim 2 P_0 \sim \tau \epsilon^{-1}/2$. 
This fact may be reflected in the difference of the shape of
the curve of $\eta$. 

In Fig. \ref{FIG2} (a), we display the results with a
low-resolution grid for $C_a=0.3$ and $\eta_0 \sim 0.01$ and 
for $C_a=0.334$ and $\eta_0 \sim 0.04$ (dotted curves).
For both cases, $\epsilon=1$. 
The growth time of $\eta$ and the maximum value of $\eta$
agree approximately with those for the high-resolution simulation.
The growth time is slightly longer for the lower resolution simulations.
This indicates that for a larger numerical dissipation, 
the growth time scale is spuriously increased. 
However, the grid resolution adopted here is high enough to
obtain a convergent result.

\subsection{The fate of unstable stars}

\begin{figure}[thb]
\begin{center}
\epsfxsize=3.in
\leavevmode
\epsffile{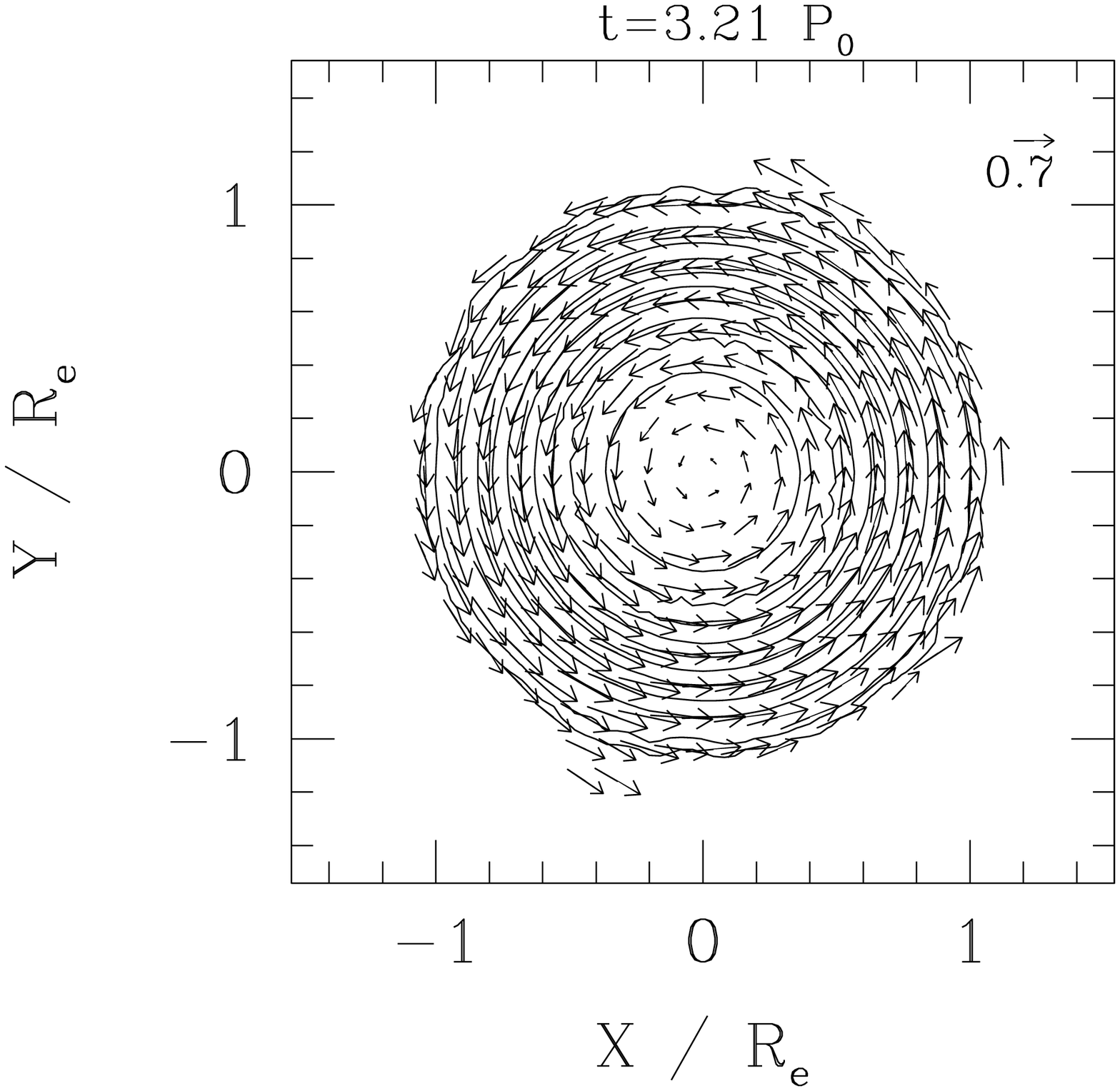}
\epsfxsize=3.in
\leavevmode
\epsffile{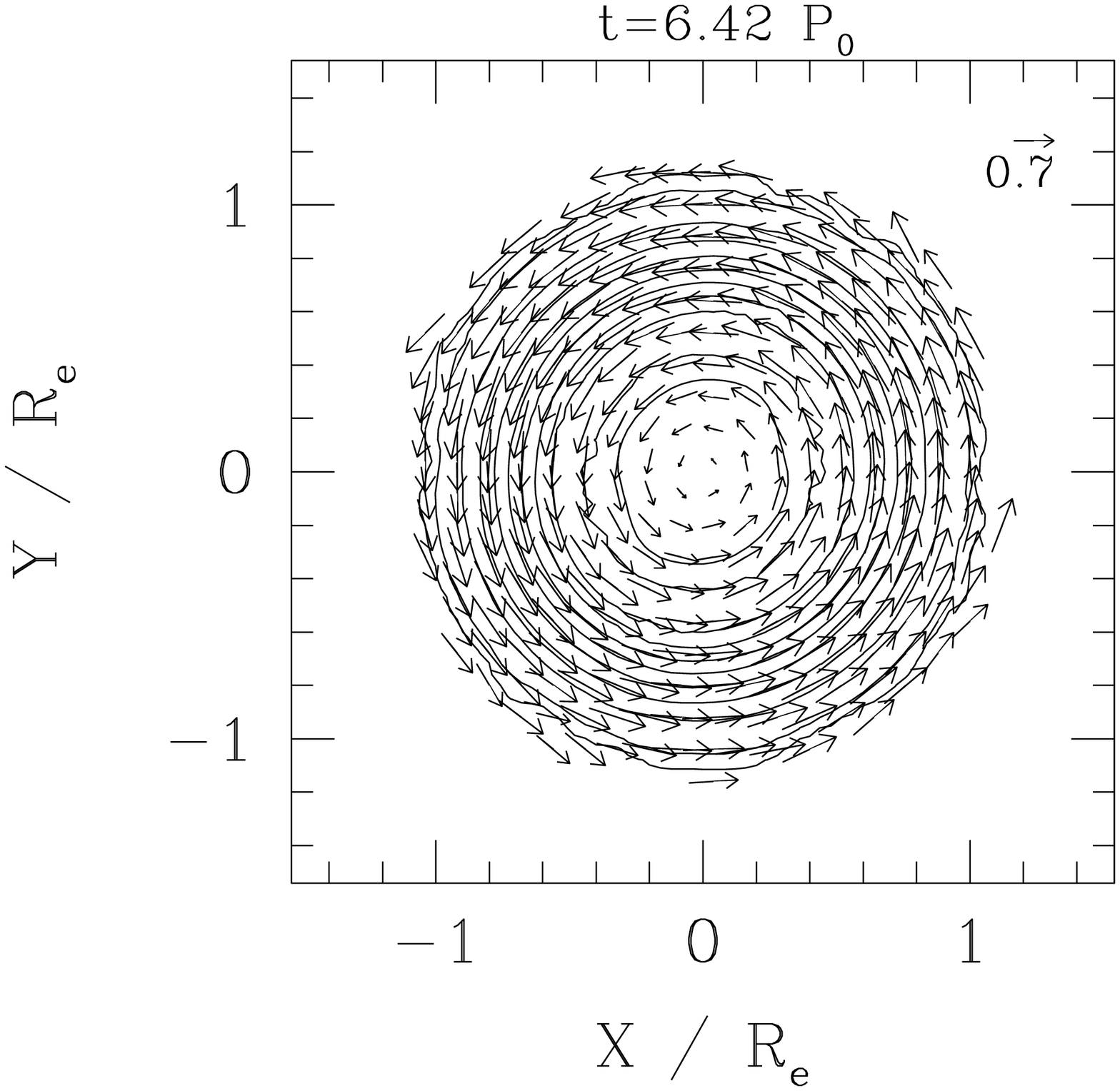}\\
\vspace*{-5mm}
\epsfxsize=3.in
\leavevmode
\epsffile{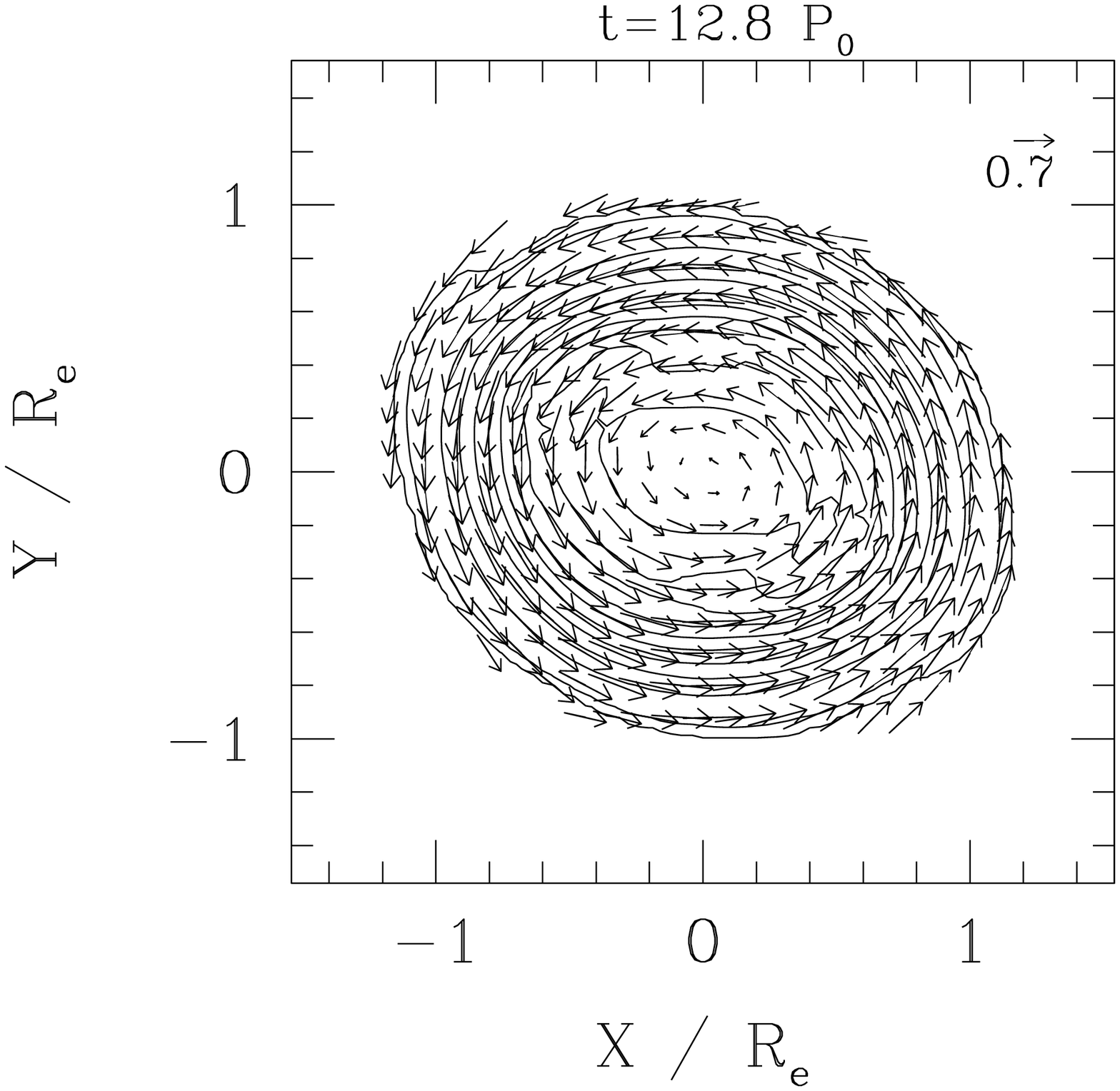}
\epsfxsize=3.in
\leavevmode
\epsffile{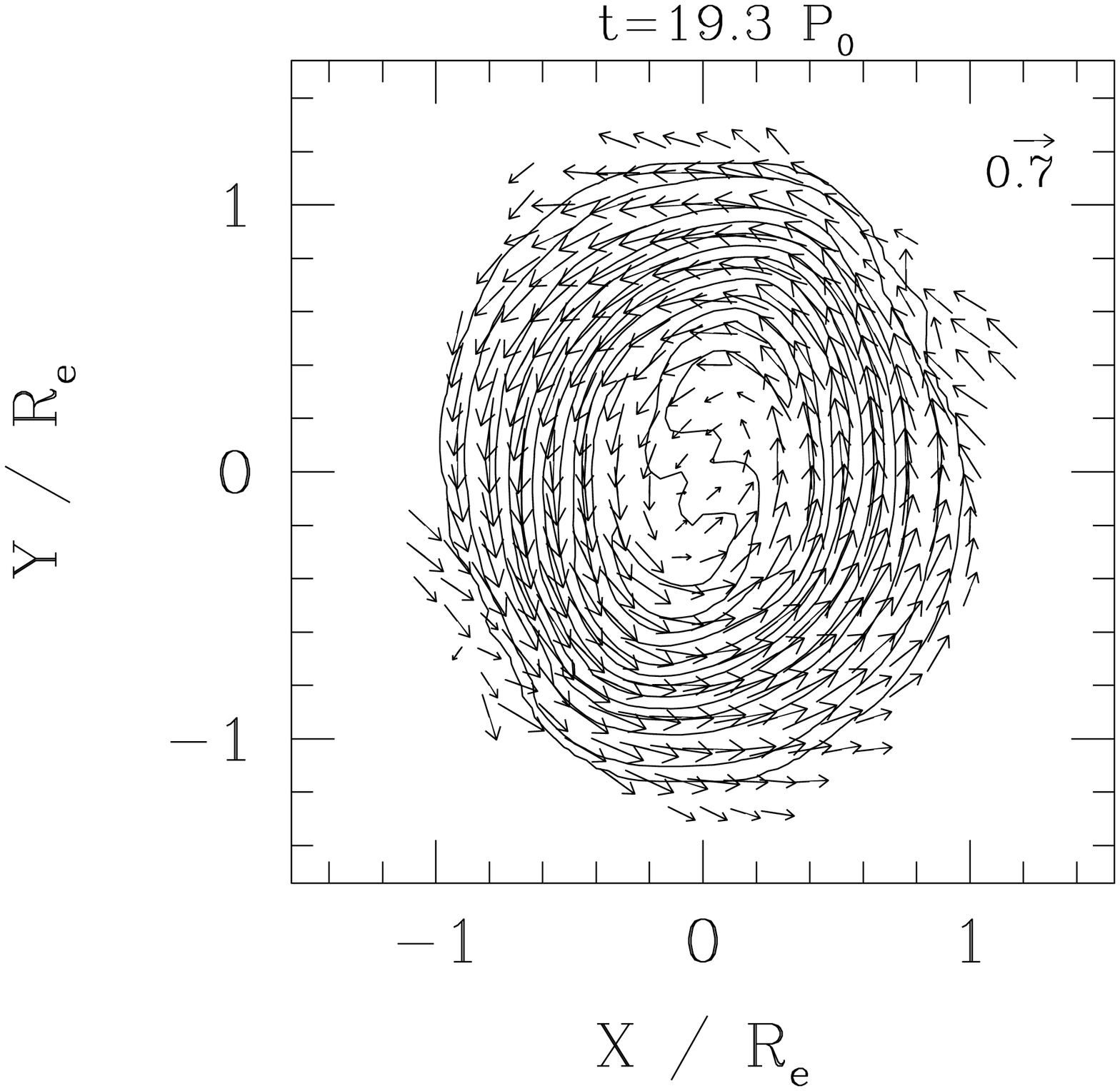}
\end{center}
\vspace*{-6mm}
\caption{
Snapshots of the density contour curves 
in the equatorial plane for $C_a=0.3$ and $\epsilon=1$.
$\eta_0 \approx 0.01$ in this case. 
The contour curves are drawn for $\rho/\rho_c=0.1j$ 
for $j=0.01, 0.1, 1, 2, \cdots, 8, 9, 9.5$, and 10. 
Vectors indicate the local velocity field $(v^x,v^y)$, and the scale 
is shown in the upper right-hand corner in units of $\rho_{c}^{1/2}R_e$.
$P_{0}$ denotes the central 
rotational period of the equilibrium configuration given at $t=0$ 
($P_0 \approx 4.674$ in units of $G=R_e=\rho_{c}=1$). 
\label{cont}}
\end{figure}

\begin{figure}[thb]
\begin{center}
\epsfxsize=3.in
\leavevmode
\epsffile{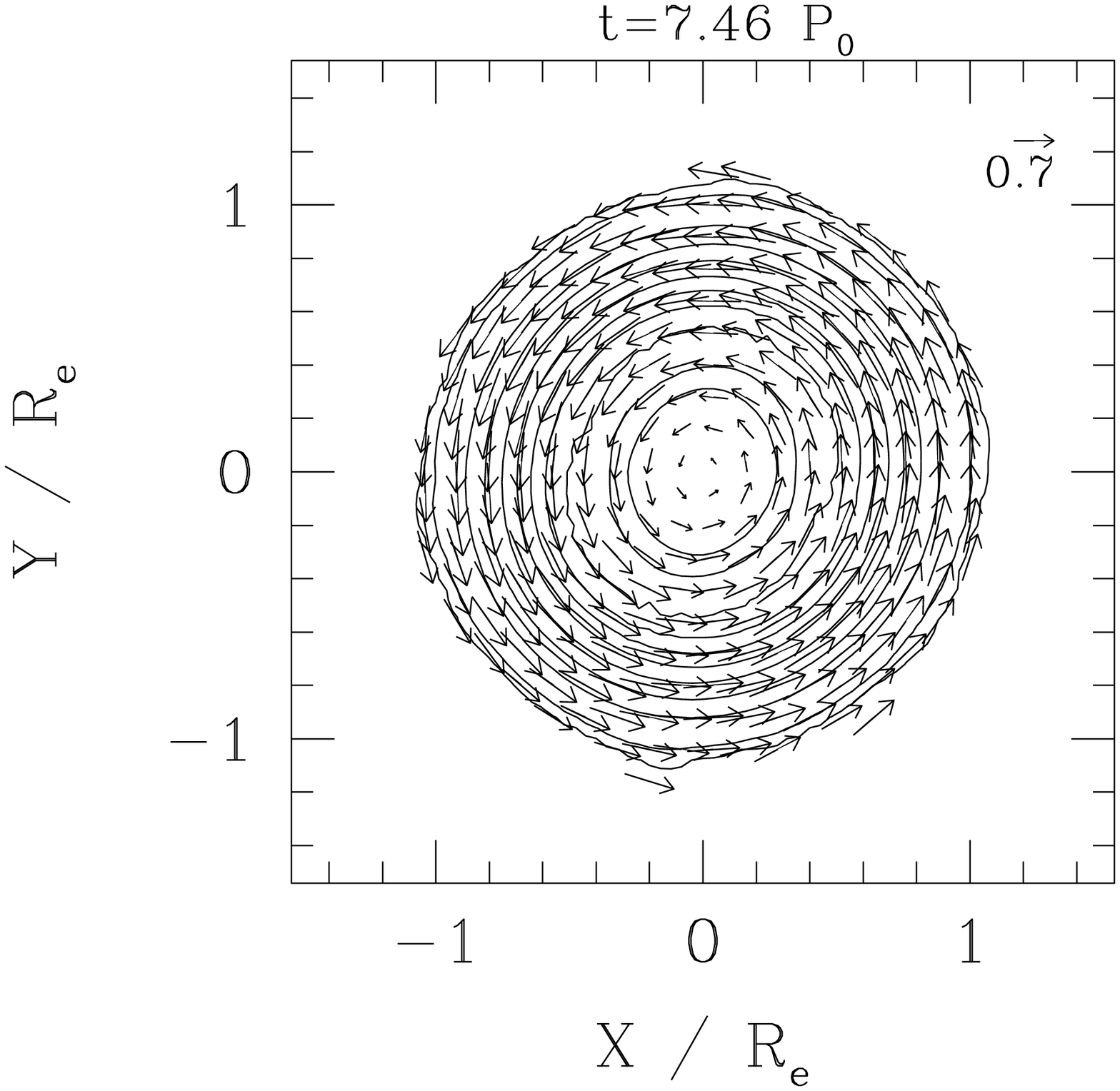}
\epsfxsize=3.in
\leavevmode
\epsffile{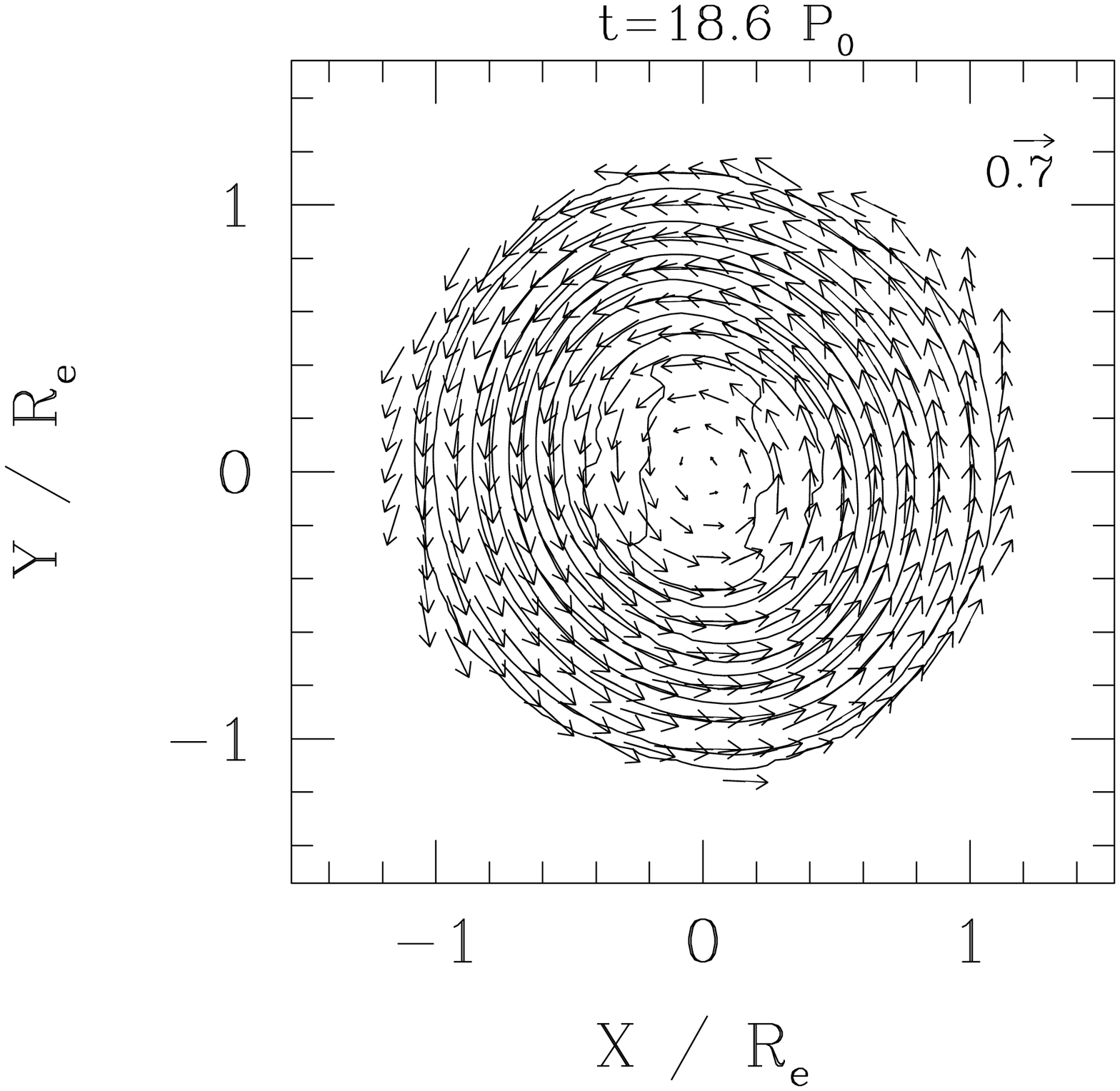}\\
\vspace*{-5mm}
\epsfxsize=3.in
\leavevmode
\epsffile{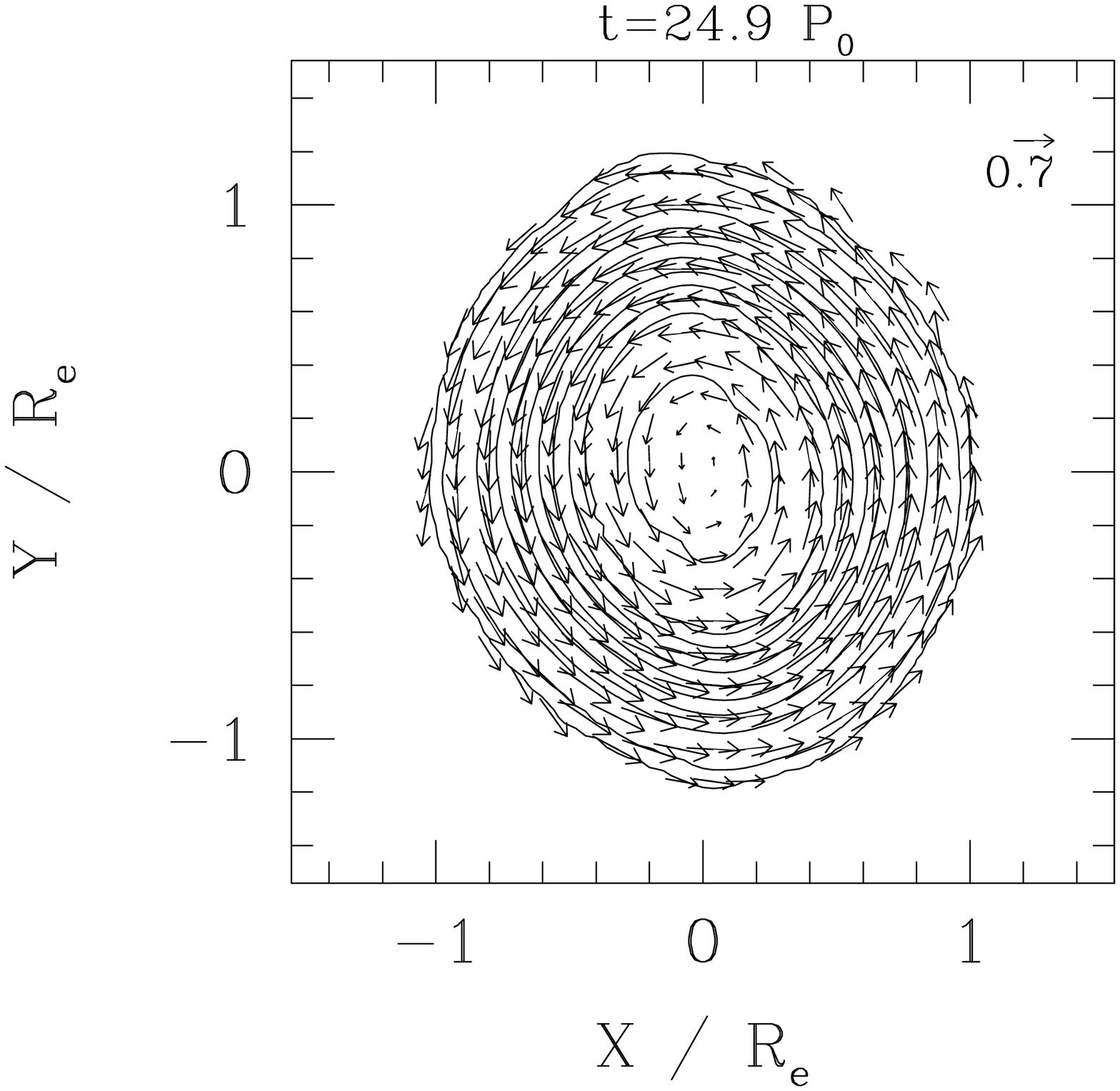}
\epsfxsize=3.in
\leavevmode
\epsffile{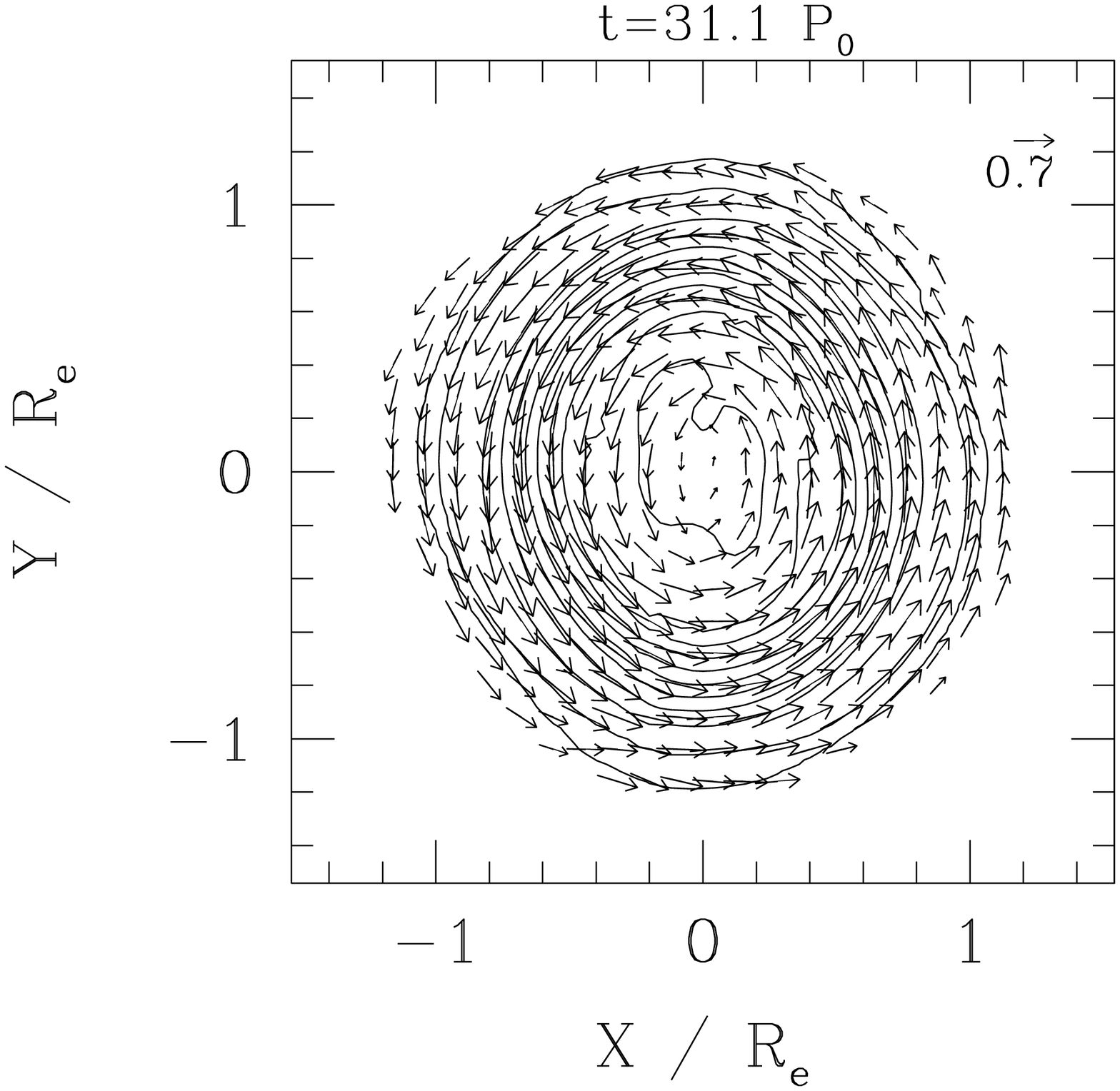}
\end{center}
\vspace*{-6mm}
\caption{
The same as Fig. \ref{cont} but for $C_a=0.334$ and $\epsilon=1$. 
$\eta_0 \approx 0.04$ in this case. 
$P_0 \approx 4.826$ in units of $G=R_e=\rho_{c}=1$. 
\label{cont2}}
\end{figure}

In Figs. \ref{cont} and \ref{cont2}, we display 
snapshots of the density contour curves and the velocity vectors 
in the equatorial plane for $C_a=0.3$, $\epsilon=1$, and
$\eta_0 \approx 0.01$, 
and for $C_a=0.334$, $\epsilon=1$, and $\eta_0 \approx 0.04$. 
With the growth of a secularly unstable and nonaxisymmetric perturbation, 
the spheroidal shape is changed to an ellipsoidal shape, and after
the value of $\eta$ reaches 0.1--0.3, 
the nonlinear growth of the perturbation terminates. 
Since the value of $\eta$ remains of order 0.1, 
the termination of the growth of the nonaxisymmetric 
deformation appears to happen when the rotating star 
reaches a quasistationary ellipsoidal state (see below for more
discussion). Here, ``quasistationary'' implies that the axial 
ratio of the ellipsoid does not change in a dynamical time scale 
and the evolution is determined by the radiation reaction time 
scale which is much longer than the dynamical time scale of system. 

The value of $\eta_{\rm max}$ does not depend strongly on the
value of $\eta_0$ [cf. Fig. \ref{FIG2}(a)], but it does on $C_a$.
In particular, the smaller the value of $C_a$,
the larger the value of $\eta_{\rm max}$ so that 
$\eta_{\rm max} \sim 0.3$, 0.2, and 0.1 for $C_a=0.3$, 0.334, and 
0.35, respectively. This feature is expected from 
a study for the incompressible fluid \cite{miller} and 
as well as from a compressible ellipsoidal model \cite{LS}. 
As the maximum value of $\eta$ is reached, 
a semi major axial ratio of the formed ellipsoid defined 
in the equatorial plane also depends on the value of $C_a$.
For $C_a=0.3$, the axial ratio of the ellipsoid 
$\approx \sqrt{(1-\eta)/(1+\eta)}$ reaches $\sim 0.75$, 
and the numerical results suggest that 
it is smaller for the larger value of $C_a$.
We note that for an incompressible star \cite{LS}, the axial ratio
reached for $\beta \sim 0.25$ should be much smaller than 0.75.
This may be a consequence that a rotating star with
larger value of $n$ is less prone to forming an ellipsoid \cite{james}. 
However, even for $\Gamma=2$, the axial ratio of the semi major axes
is fairly large for $C_a \alt 0.35$, and therefore, 
the ellipsoid will subsequently 
emit gravitational waves of large amplitude and 
dissipate energy and angular momentum due to a longterm 
emission of gravitational waves (cf. Sec. III C).

\begin{figure}[thb]
\begin{center}
\epsfxsize=3.in
\leavevmode
(a) \epsffile{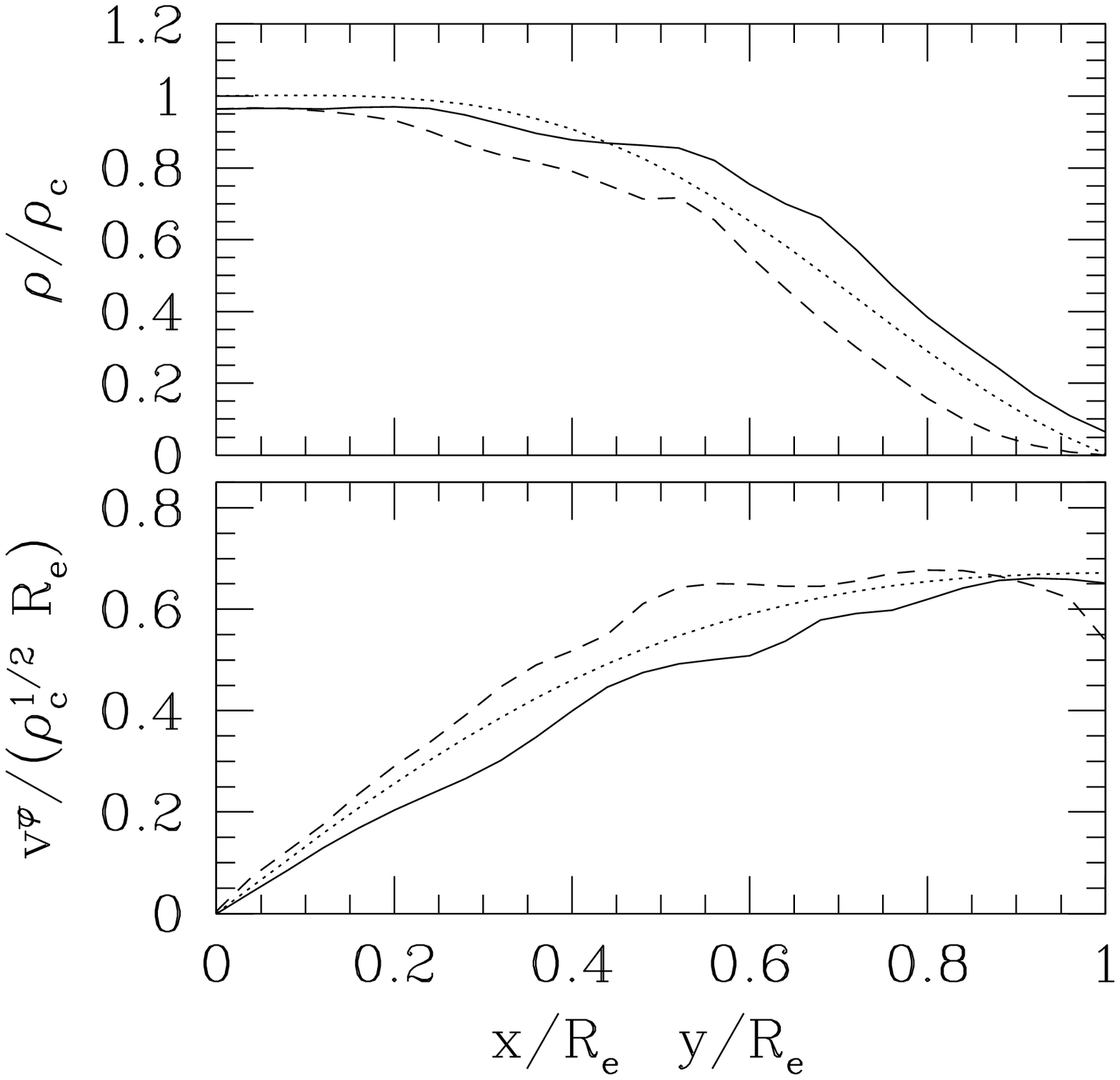}
\epsfxsize=3.in
\leavevmode
~~(b)
\epsffile{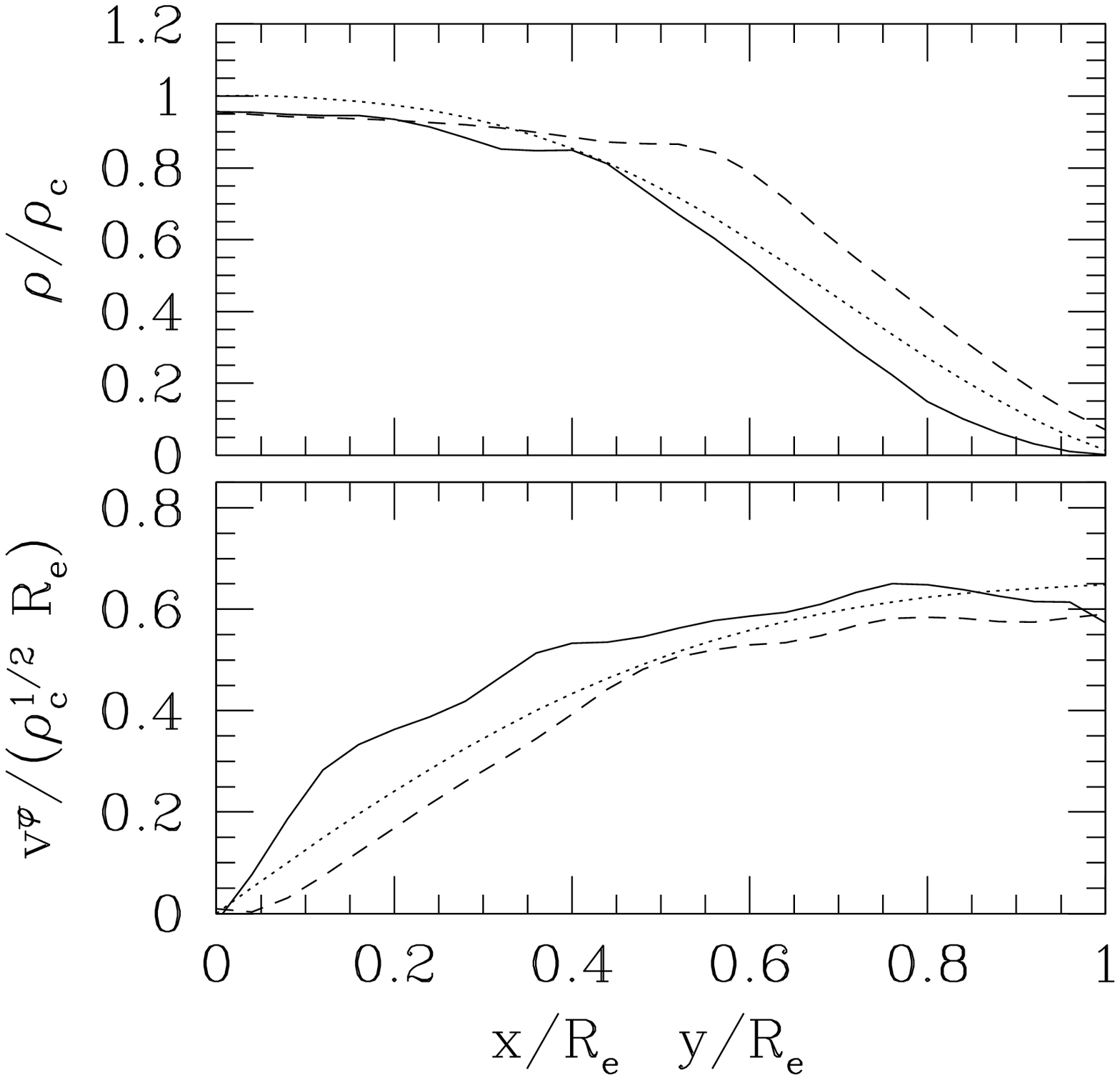}
\end{center}
\vspace*{-2mm}
\caption{(a) The density and angular velocity profile along
$x$ and $y$ axes at 
$t=0$ (dotted curves) and $t=12.8 P_0$ (solid and dashed curves)
for $C_a=0.3$ and $\epsilon=1$. The solid and dashed curves denote
the profiles along $x$ and $y$ axes, respectively.
(b) The same as (a) but for $C_a=0.334$ and $t=24.9 P_0$. 
\label{omega}}
\end{figure}

In Fig. \ref{omega}, we show the profiles of 
the density and angular velocity along the $x$ and $y$ axes at
selected time slices. The contour curves and the velocity vectors
at the corresponding time steps are shown in the third
panels of Figs. \ref{cont} and \ref{cont2}. In other words, 
Figures \ref{omega}(a) refers to the profile 
at $t/P_0=12.8$ for $C_a=0.3$ and $\epsilon=1$, and 
Fig. \ref{omega}(b) is at $t/P_0=24.9$ 
for $C_a=0.334$ and $\epsilon=1$ with $\eta_0 \approx 0.04$. 
In each figure, the density and the angular velocity 
at $t/P_0=0$ (dotted curves) are shown together. 
Around the origin, the density profile does not change
significantly even after the formation of the ellipsoid.
(Note that the central density decreases 
with time partly due to a numerical dissipation or diffusion.) 
However, around $\varpi/R_e \sim 0.5$--0.6, a density peak 
along the semi major axis is formed irrespective of the
initial condition. Also, it is found that a nonaxisymmetry
is enhanced around $\varpi/R_e > 0.2$. 
These features found in the density profiles reflect the
initial perturbation profile [cf. Fig. 1(a)].
After the formation of the ellipsoid, 
differential rotation is also enhanced as the result of the growth of 
the nonaxisymmetric density perturbation. 
It is interesting to note that a high-density and 
nonaxisymmetric peak rotates with a smaller rotational speed than
that for the unperturbed axisymmetric star.
This seems to be reasonable because the secularly unstable 
mode has an angular velocity smaller than
the rotational angular velocity of the unperturbed equilibrium state.

\begin{figure}[thb]
\begin{center}
\epsfxsize=3.in
\leavevmode
\epsffile{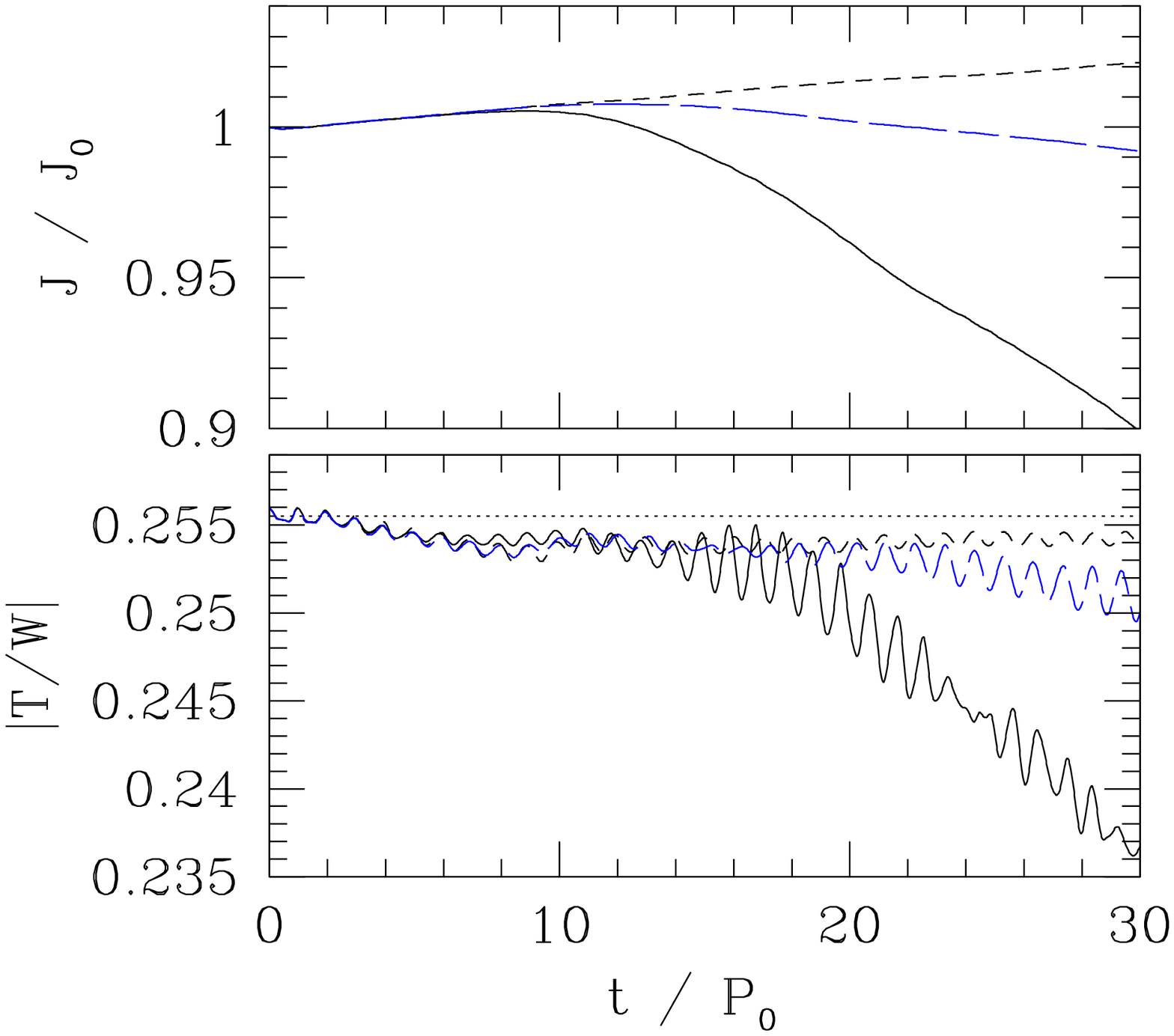}
\end{center}
\vspace*{-5mm}
\caption{Evolution of $J$ and $\beta=T/|W|$ for $C_a=0.3$, and
$\epsilon=1$ (solid curve), 0.7 (long dashed curve), and 0 (dashed curve).
$J_0$ denotes the initial value of $J$. 
\label{beta}}
\end{figure}

In Fig. \ref{beta}, the evolution of the total angular momentum $J$ and 
$\beta$ are shown for $C_a=0.3$, and 
$\epsilon=1$ (solid curve), 0.7 (long dashed curve), and 0 (dashed curve). 
Even for $\epsilon=0$ for which the system should be almost stationary 
throughout the simulation, 
$J$ increases by $\sim 2\%$ and $\beta$ decreases by $\sim 1\%$ in $30 P_0$. 
These are the typical truncation error 
with grid of $141\times 141 \times 141$. For $\epsilon > 0$, 
during the exponential growth of the secularly unstable mode, 
$J$ and $\beta$ decrease at most 1\% by gravitational wave emission.
(Compare the curves with $\epsilon=0$ and $\epsilon \not=0$.
The difference between two values can be regarded as loss by
the gravitational radiation.) 
Thus, the formed ellipsoid initially has slightly smaller values of
$J$ and $\beta$ than those of the secularly unstable axisymmetric
star in equilibrium. 
After the formation of a quasistationary ellipsoid, these values 
start decreasing significantly 
due to dissipation by gravitational waves. 
Figure \ref{beta} indicates that when the ellipsoid relaxes to
a final stationary state that does not emit gravitational waves, 
the values of $J$ and $\beta$ will be much smaller
than the initial values.  

The final state after a sufficient emission of gravitational waves 
is uncertain. For an incompressible fluid star, it will be a Dedekind 
ellipsoid which is static in shape and has
only internal motion and no pattern rotation if it is observed 
in the inertial frame. 
However, for $n \agt 0.8$, such configuration may not 
exist because of the following facts:
(i) the uniformly rotating ellipsoid (Jacobi-like ellipsoid)
does not exist for $n > 0.808$ \cite{james} and
(ii) the Dedekind theorem for 
the incompressible fluid tells that the absence of the
Jacobi ellipsoid implies the absence of the Dedekind ellipsoid \cite{CH69}.
Namely, if this theorem holds for $n \not=0$, no Dedekind-like 
configuration would exist for $n > 0.808$. 
Lai and Shapiro expect that the final state 
may be a Dedekind-like ellipsoid even for $n > 0.8$ \cite{LS}. 
The other possibility is that the unstable star 
does not relax to a Dedekind-like ellipsoid but
to a (nearly) axisymmetric and secularly stable rotating star
after sufficient emission of gravitational waves.
The results of the present simulation 
indicate that the latter is more likely because of the following 
reasons. 

For the incompressible case, a secularly unstable Maclaurin spheroid 
changes to a Dedekind ellipsoid by emission of gravitational 
waves conserving the vorticity \cite{LS}. Throughout this transition, 
the energy and the angular momentum are decreased at most by a few percents 
from their initial values \cite{LS}. Furthermore, for
$\beta \sim 0.25$, the axial ratio in the equatorial plane becomes
very small $\ll 0.75$. In the case of $C_a=0.3$, $\eta_0 \sim 0.01$, and
$\epsilon=1$, until the end of the simulation 
at $t\sim 30P_0$, the energy and the angular momentum are 
dissipated by gravitational waves by $\sim 5\%$ and $\sim 10\%$, 
respectively. However, even at $t \sim 30 P_0$, 
the rotating stars still do not settle into 
a highly deformed Dedekind-like ellipsoid but to a 
moderately deformed quasistationary (Riemann-type) ellipsoid
which emits gravitational waves of a moderately
large amplitude (cf. Sec. III C). 
The results of the simulation also indicate 
that the gravitational wave emission 
does not enforce the ellipsoid to a stationary state immediately,
although the final outcome should be an object that does not
emit gravitational waves. Thus, the ellipsoid will 
subsequently emit gravitational waves for a long time 
to dissipate the energy and the angular momentum by $\agt 10\%$.
Indeed, Fig. \ref{beta} suggests that $\beta$ will decrease much
below the initial value. Therefore, in the present case,
the final state after a longterm dissipation by gravitational waves
will not be a highly deformed Dedekind-like ellipsoid, 
but a nearly axisymmetric spheroid of $\beta \ll 0.25$. 

As mentioned above, the uncertainty on the
existence of the Dedekind ellipsoid for 
compressible stars may be a key issue in understanding the final 
state after a longterm emission of gravitational waves.
To investigate whether a compressible Dedekind equilibrium exists or not,
a specially designed implementation 
is necessary. Unfortunately, systematic numerical computation of
the compressible Dedekind ellipsoid has not been succeeded yet  
because of the technical difficulty (but see \cite{UE}).
A robust numerical implementation is required 
to clarify the possible final state after a longterm 
dissipation by gravitational waves.

\subsection{Gravitational waves}

\begin{figure}[thb]
\begin{center}
\epsfxsize=3.in
\leavevmode
(a)\epsffile{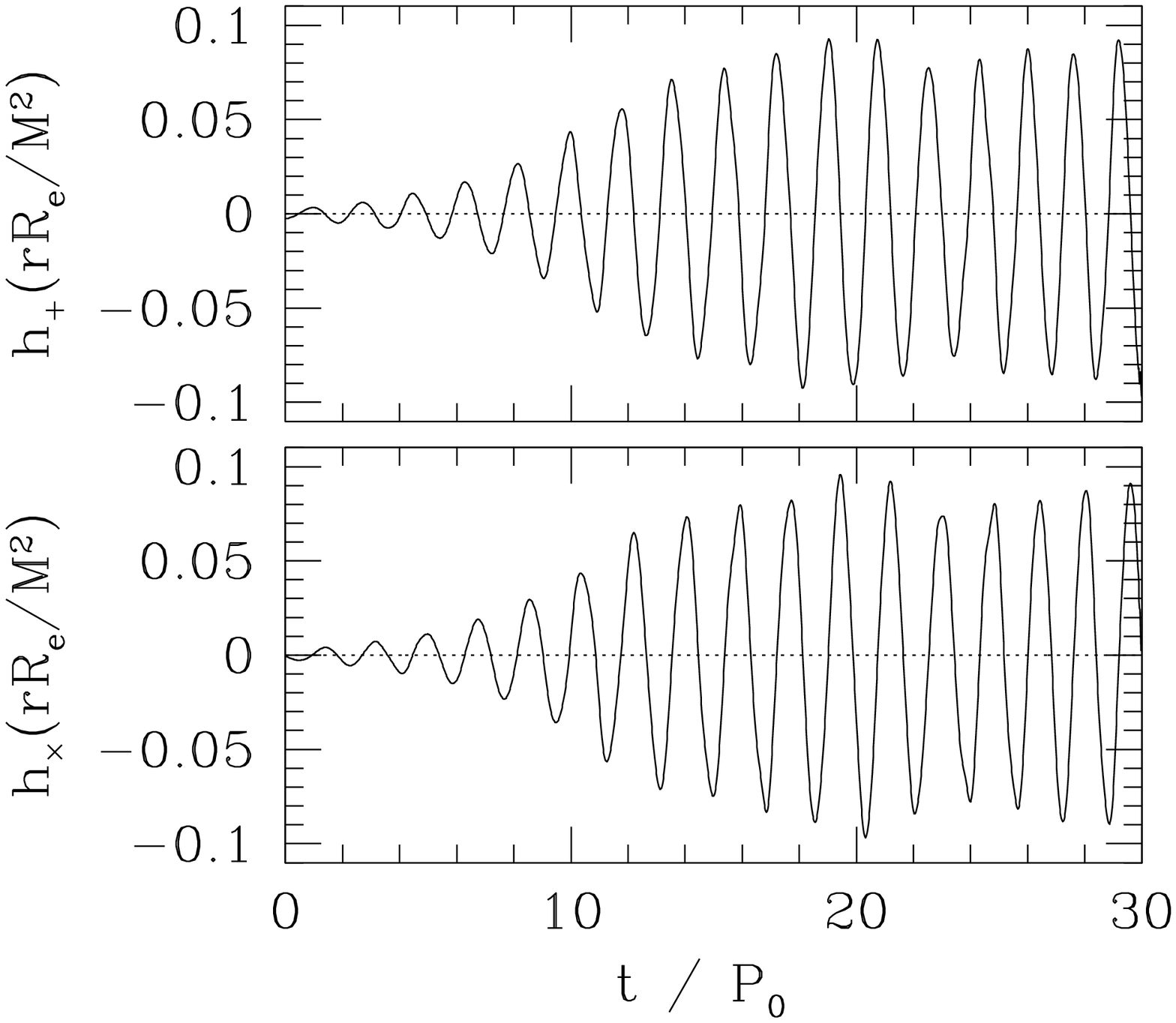}
\epsfxsize=3.in
\leavevmode
~~(b)\epsffile{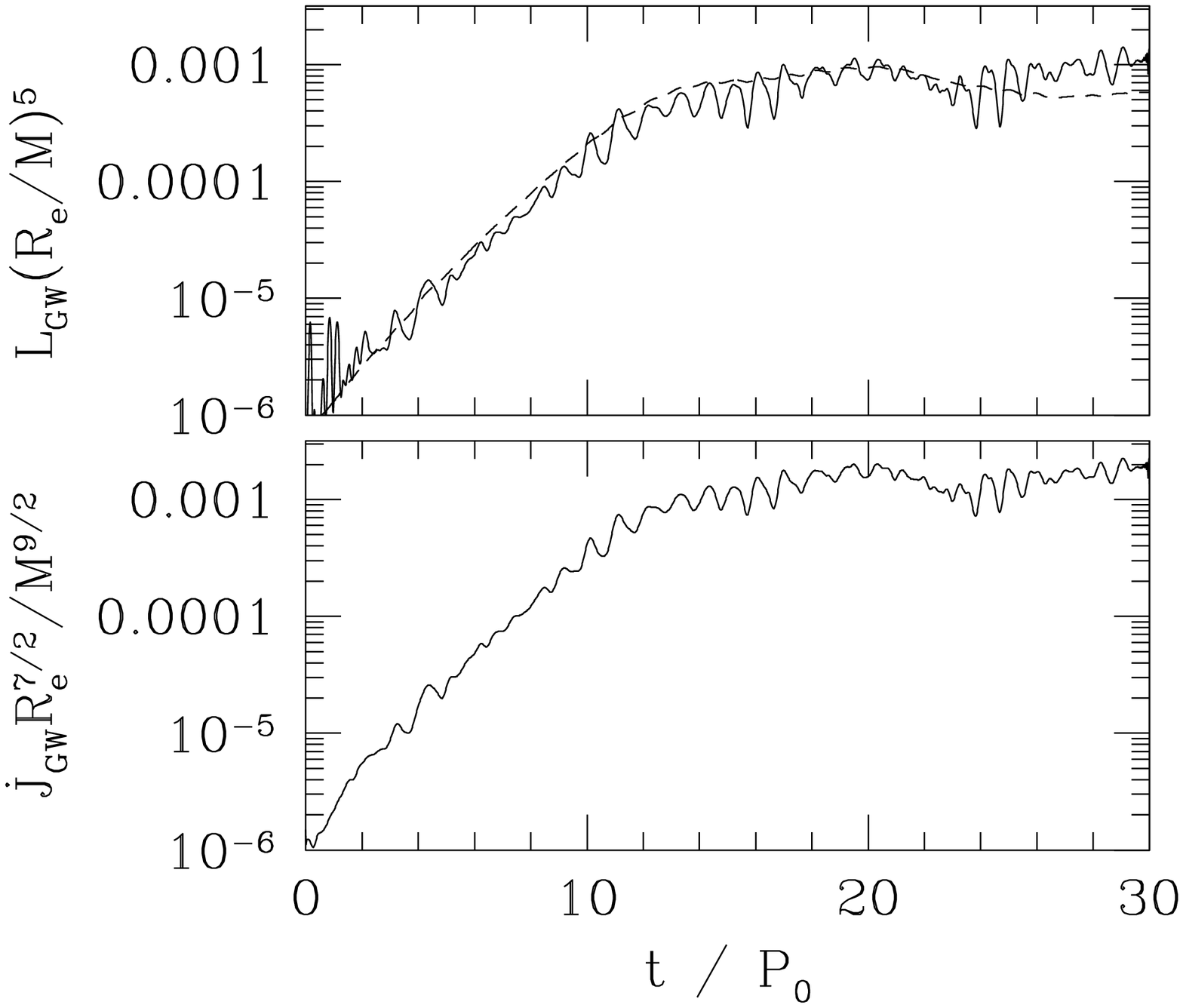}
\end{center}
\caption{(a) Gravitational waveforms for $C_a=0.3$ with
$\epsilon=1$. The upper and lower figures show 
the $+$ and $\times$ modes, respectively.
(b)  The same as (a) but for luminosity and angular momentum flux. 
The solid and dashed curves in the upper figure denote $L_{\rm GW}$ and 
$L_{\rm GW1}$, respectively. 
\label{GW1}}
\end{figure}

In Fig. \ref{GW1}, the gravitational waveforms, luminosity, and
angular momentum flux for $C_a=0.3$ with $\epsilon=1$ are shown. 
In the early stage in which $\eta$ increases exponentially, 
the amplitude and the fluxes of gravitational waves 
are increased in an exponential manner.
After the end of the nonlinear growth of the nonaxisymmetric
perturbation, they relax approximately to constant values.
(The reason that they do not exactly settle into constants is that 
modes other than the secularly unstable mode are excited.)
This global feature of gravitational waveforms 
is independent of the initial value of $C_a$,
although the amplitude is different depending on the value of $\eta$. 

When the secularly unstable mode of the angular velocity $\sigma$ is
the only one present mode, 
the luminosity of gravitational waves should be equal to 
\beq
L_{\rm GW1}={1 \over 10} \sigma^6 \eta^2 I_{\varpi\varpi}^2. 
\eeq
To confirm that the amplitude of gravitational waves is increased mainly 
by the growth of the secularly unstable mode, we compare the 
energy flux $L_{\rm GW1}$ with $L_{\rm GW}$. 
In Fig. \ref{GW1}(b), the evolution of $L_{\rm GW1}$ is shown 
together (dashed curve). We see that the luminosities computed 
by two methods agree approximately, although in the curve corresponding 
to $L_{\rm GW}$ shows oscillations. This is due to 
the fact that modes other than the secularly unstable one are excited. 

From a break of the luminosity curve at $t \sim 12P_0$, 
we can approximately identify the end-point of the 
exponentially growing stage of the secularly unstable perturbation. 
Until this time, the energy and the angular momentum
carried away by gravitational waves are only by $\alt 1\%$ of the 
total energy and angular momentum for all the models. 
This implies that the quasistationary ellipsoid formed 
at the end of the growth of the perturbation
has only slightly smaller energy and
angular momentum than those of the secularly unstable and
axisymmetric rotating star. In contrast to the incompressible case
\cite{LS}, most of the rotational kinetic energy and the angular momentum
are dissipated by a subsequent emission of gravitational waves
from a formed ellipsoid as discussed in Sec. III B. 
In the following, let us estimate the accumulated gravitational
wave amplitude assuming that the ellipsoid is a protoneutron star. 

For the luminosity $L_{\rm GW} = 0.001 \alpha (M/R_e)^5$ and
the kinetic energy $T = \zeta (M^2/R_e)$ 
where $\alpha$ and $\zeta$ are parameters, the emission time scale 
of gravitational waves of a protoneutron star can be estimated as
\beqn
\tau_{\rm evo} \equiv {T \over L_{\rm GW}} &=& 10^3 \alpha^{-1} \zeta 
\biggl({R_e \over M}\biggr)^4 M \nonumber \\ 
&\approx& 12 \alpha^{-1} \biggl({\zeta \over 0.2}\biggr) 
\biggl({R_e \over 20~{\rm km}}\biggr)^4 
\biggl({M \over 1.4M_{\odot}}\biggr)^{-3}~{\rm sec}. 
\eeqn
Here, $\alpha \approx (\eta/0.3)^2 \alt 1$ and $\zeta \sim 0.2$ 
for $\beta = 0.2$--0.25, respectively. 
The value of $R_e$ will depend strongly on
the rotational profile of a progenitor of supernova stellar collapse
(e.g., \cite{ZM}). In the present case, we assume that
protoneutron stars are rapidly rotating with $\beta > 0.2$. In such case, 
the equatorial radius would be larger than a canonical radius of
neutron stars $\sim 10$--15 km because of the strong centrifugal force.
For a sufficiently large initial value of $\beta \agt 0.01$
and for a small value of $\hat A \alt 1/2$, the outcome of 
stellar collapse could be an oscillating star of subnuclear density
and of a high value of $\beta$.
In such case, $R_e$ may be as large as $\sim 100$ km. 

The characteristic frequency of gravitational waves is denoted as 
\beqn
f \approx 460 \biggl({\sigma \over 0.6 \rho_c^{1/2}}\biggr)
\biggl({R_e \over 20~{\rm km}}\biggr)^{-3/2}
\biggl({M \over 1.4M_{\odot}}\biggr)^{1/2}~{\rm Hz}, 
\eeqn
where $\sigma$ is the angular velocity of the nonaxisymmetric 
perturbation in units of $\rho_c^{1/2}$, where $\rho_c$ is the initial 
central density. $\sigma/\rho_c^{1/2}$ depends strongly on 
the value of $\beta$, but the value of $\sigma$
is always smaller than $\Omega_0$ by a factor of $\agt 2$. 
This is a very important feature for the detection of gravitational
waves by kilometer-size interferometers 
for which a sensitivity is 
the best in the frequency band around a few 100 Hz $\ll \Omega_0/\pi$. 

Assuming that the nonaxisymmetric perturbation would not be dissipated 
by viscosities or magnetic fields on the emission time scale of gravitational 
waves \cite{BSS}, the accumulated cycles of gravitational wave-train $N$ 
are estimated as 
\beqn
N \equiv f\tau_{\rm evo} & \approx & 5.5 \times 10^3 \alpha^{-1}
\biggl({\zeta \over 0.2}\biggr) 
\biggl({\sigma \over 0.6 \rho_c^{1/2}}\biggr) 
\biggl({R_e \over 20~{\rm km}}\biggr)^{5/2}
\biggl({M \over 1.4M_{\odot}}\biggr)^{-5/2}. 
\eeqn
In reality, the characteristic frequency will be changed 
as gravitational waves are emitted, and hence, the expression 
of $N$ which denotes the accumulated cycle
for a given frequency should be regarded as the maximum value. 
However, the order of magnitude for the real value of $N$ 
will be identical. 

The effective amplitude of gravitational waves is defined by 
$h_{\rm eff} \equiv N^{1/2}h$ \cite{Kip,LS,LL1} 
where $h$ denotes the characteristic amplitude 
of periodic gravitational waves. 
According to the numerical results,
\beqn
h = 0.1 \bar h {M^2 \over r R_e} \approx 7.0 \times 10^{-24} \bar h
\biggl({M \over 1.4M_{\odot}}\biggr)^2 
\biggl({R_e \over 20~{\rm km}}\biggr)^{-1} 
\biggl({r \over 100~{\rm Mpc}}\biggr)^{-1}, 
\eeqn
where $\bar h \propto \eta$ and $\sim \alpha^{1/2} \alt 1$. 
Using this relation, 
\beqn
h_{\rm eff}&\approx&
5.2 \times 10^{-22} \alpha^{-1/2} \bar h 
\biggl({\zeta \over 0.2}\biggr)^{1/2}
\biggl({\sigma \over 0.6 \rho_c^{1/2}}\biggr)^{1/2}
\biggl({R_e \over 20~{\rm km}}\biggr)^{1/4} 
\biggl({M \over 1.4M_{\odot}}\biggr)^{3/4}
\biggl({r \over 100~{\rm Mpc}}\biggr)^{-1}. \label{eqh}
\eeqn
Here, $\alpha^{-1/2} \bar h \sim 1$ irrespective of $\eta$, and 
thus, $h_{\rm eff}$ is approximately
independent of $\eta$. This property is qualitatively explained as follows: 
As gravitational waves are emitted, the value of $\eta$ will be 
decreased. This implies that the amplitude of gravitational waves $h$ 
and luminosity are decreased.
However, the time scale of gravitational radiation reaction
is increased with decreasing the luminosity. These two effects 
cancel each other as that $h_{\rm eff}$ is approximately
independent of $\eta$. 
Thus, Eq. (\ref{eqh}) may be used for any value of $\eta$, i.e.,
for any initial value of $\beta$. (Remember that 
the maximum value of $\eta$ is smaller for smaller values of $\beta$.) 

Equation (\ref{eqh}) shows that 
$h_{\rm eff}$ will be larger than $10^{-21}$ at a distance 
$r \alt 50$ Mpc with the frequency of about 500 Hz 
for $R_e \sim 20$ km and $M \approx 1.4M_{\odot}$. 
As gravitational waves are emitted, the rotational velocity
of the ellipsoid will be decreased, and hence, 
the frequency of gravitational waves and the value of $\zeta$
will be also decreased. However, 
$h_{\rm eff}$ is proportional to $\sigma^{1/2}$ and $\zeta^{1/2}$ 
for given values of mass and radius, and thus, 
the dependence of $h_{\rm eff}$ on these two parameters is not very
strong \cite{LS}. Therefore, 
the value of $h_{\rm eff}$ could be $\agt 10^{-21}$ 
at the frequency at a distance of $\sim 10$ Mpc 
in a wide range between $\sim 10$ Hz and $\sim 500$ Hz 
for $R_e \sim 20$ km and $M \approx 1.4M_{\odot}$. 
(This amplitude agrees approximately with that derived in \cite{LS}.) 
This implies that 
gravitational waves from protoneutron stars of a high value of
$\beta$ and of mass $\sim 1.4 M_{\odot}$ and radius $\sim 20$ km at a 
distance of $\sim 10$ Mpc are likely to be broadband and quasiperiodic 
sources for laser interferometric detectors such as LIGO \cite{Kip2}. 

If other dissipation or transport processes of the angular momentum, 
e.g., by viscosity or magnetic field,  is present, 
growth of the nonaxisymmetric perturbation
in a protoneutron star may be suppressed.
A detailed discussion about the viscous effects is found in
\cite{LS}, in which the authors indicate that the viscous effect
will be negligible. 
On the other hand, effects due to magnetic fields have not been
investigated in detail. In \cite{BSS}, the authors estimate
the time scale of magnetic braking which transports the angular momentum
in a differentially rotating star outward in 
$\sim 10(B/10^{12}{\rm Gauss})$ sec where $B$ is a 
strength of the magnetic field. For a canonical value of
$B = 10^{12}$ Gauss, the time scale is comparable to $\tau_{\rm evo}$.
Therefore, in the presence of the magnetic fields, 
a rotating ellipsoidal star may not evolve simply due to 
gravitational radiation reaction. This point is not clear at present. 

\section{Summary}

We have numerically studied the growth of a secularly unstable 
bar-mode induced by gravitational radiation for rapidly rotating 
protoneutron stars of $\beta$ between $\sim 0.2$ and $\sim 0.25$ 
with the $\Gamma=2$ polytropic equation of state. 
Numerical simulations were performed in a (0+2.5) post-Newtonian 
framework in which the effect of gravitational radiation reaction 
is incorporated in addition to Newtonian gravity. 
In our simulations, the amplitude of the secularly unstable 
bar-mode is driven to a nonlinear stage in which the deformation
parameter becomes 0.1--0.3 depending on the initial value of
$\beta$. The outcome is a moderately deformed
ellipsoid of the semi major axial ratio $\agt 0.75$, in contrast to
the incompressible case, in which it becomes much smaller than 0.75
for $\beta \sim 0.25$. 
The final values of the deformation parameter $\eta$ 
and of the axial ratio do not depend 
strongly on the radiation reaction force. However, 
they depend on the initial value of $\beta$, and are smaller for the 
smaller value of $\beta$, as in the incompressible and 
rigidly rotating case \cite{miller,LS}. 
The growth of the nonlinear perturbation appears to be terminated 
because the formed ellipsoid is quasistationary. The total 
energy and angular momentum dissipated by gravitational waves until the 
formation of the ellipsoid is $\sim 1\%$ of the initial values. Thus, 
the formed ellipsoid initially has only slightly smaller energy and
angular momentum than those for the axisymmetric rotating star 
adopted at $t=0$. 

In contrast to the incompressible and rigidly rotating case,
the formed quasistationary ellipsoid is not likely to be
a Dedekind-like stationary ellipsoid. Namely, their pattern 
rotates globally, and hence, gravitational waves are emitted. 
The simulations indicate that the gravitational wave emission does not 
enforce the ellipsoid to a stationary state immediately, 
although the final state should be such a state that does not 
emit gravitational waves. Thus, the ellipsoid will 
subsequently emit gravitational waves for a long time 
to dissipate the energy and the angular momentum significantly. 
This implies that the final state may not be a highly deformed 
Dedekind-like ellipsoid but a nearly axisymmetric spheroid. 
Also, this indicates that in the absence of other dissipative 
mechanisms, a rapidly rotating protoneutron star may be 
a stronger emitter than that considered before \cite{LS}, 
because the energy and the angular momentum will be dissipated 
by gravitational waves by $\agt 10\%$, not a few \%, until 
formation of a stationary star. 

The growth time of $\eta$ (denoted by $\tau$) is evaluated
using numerical data sets and compared with an analytic formula
$\tau_{\rm anal}$ for incompressible 
and rigidly rotating stars. It is found that the value of $\tau$ 
is systematically smaller than that derived from $\tau_{\rm anal}$. 
Plausible explanations for this difference are either of the following facts: 
(i) the analytic formula is valid only for rigidly rotating stars and, 
hence, in differentially rotating stars, $\tau$ becomes systematically 
shorter than $\tau_{\rm anal}$;
(ii) the analytical formula is valid only for incompressible
stars and for compressible stars, $\tau$ becomes systematically 
shorter than $\tau_{\rm anal}$. 
To clarify the plausible reason, a linear perturbative analysis 
including gravitational radiation reaction will be the 
best approach. 

An ellipsoidal neutron star formed after the growth of
a secularly unstable mode of 
a rapidly rotating progenitor has a fairly large ellipticity and a
pattern rotation. The angular velocity of the pattern rotation
is much smaller than that of a axisymmetrically rotating
equilibrium star initially given. 
Thus, the formed ellipsoid will be a strong emitter of gravitational
waves of fairly low frequency, less than several hundreds Hz. 
Using a simple analysis, an effective amplitude 
of gravitational waves is evaluated. It is found that 
for a protoneutron star of the equatorial radius 
$R_e \sim 20$ km, mass $M \approx 1.4M_{\odot}$ and 
the initial value of $\beta \sim 0.25$, 
the effective amplitude $h_{\rm eff}$ 
will be larger than $10^{-21}$ at a distance 
$r \sim 10$ Mpc in a wide frequency range between
$\sim 10$ and $\sim 500$ Hz. Such gravitational waves
from protoneutron stars are likely to be sources for 
laser interferometric detectors such as LIGO \cite{Kip2}.
In the presence of magnetic fields of magnitude $\agt 10^{12}$ Gauss, 
magnetic braking may affect the evolution of 
differentially rotating and ellipsoidal protoneutron 
stars significantly. In such case, the evolution for the 
ellipsoidal protoneutron star is not yet clear at present, and
thus, a further study is necessary. 

{\em Note added in proof}: Soon after this paper was submitted, 
a paper by Ou, Tohline, and Lindblom \cite{add}, which 
also studies the secular instability driven by gravitational 
radiation, was posted in astro-ph. They performed a (0+2.5) post-Newtonian 
simulation for a rapidly rotating polytropic star with $\Gamma=3$ including
a radiation reaction term in essentially the same approximate 
manner as that adopted in \cite{LTV}. 
They find the similar result to that found in our present paper 
for the growth of the secular instability to form an ellipsoid. 
The formed ellipsoid has a pattern rotation of small velocity, 
and thus, is not the Dedekind but the slowly rotating Riemann-type 
ellipsoid. On the other hand, they chose $\Gamma=3$ that is larger 
than our choice ($\Gamma=2$). As a reasonable result, they 
find that the axial ratio of the formed ellipsoid is slightly smaller
($\sim 0.5$) than that found in our work. 
The radiation reaction formalism we adopt 
can be used for any problem as long as the system adiabatically 
evolves due to the back reaction. In contrast, 
the radiation reaction formalism they adopt can be in principle used to follow 
the evolution of the secular instability only from the linear to weakly 
nonlinear stages of a monochromatic frequency of gravitational waves 
in contrast to that adopted in this paper. 
This is because their basic equations are formulated 
assuming that only one nonaxisymmetric mode exists (i.e., assuming that 
the oscillation frequency of the quadrupole moment is monochromatic). 
Thus, from their simulation, it seems to be difficult to accurately 
determine the final outcome after the growth of 
the secular instability saturates in which the frequency of
quadrupole moment and gravitational waves is not 
monochromatic but gradually changed. 
On the other hand, their method may be robust to study the early growth
of the secular instability in which the frequency of gravitational waves 
is monochromatic.

\section*{Acknowledgments}

We thank Y. Eriguchi for valuable comments and discussion, and 
L. Rezzolla for helpful comments on the presentation. 
Numerical simulations were performed on FACOM VPP5000 in the 
data processing center of National Astronomical Observatory of Japan. 
This work was in part supported by Japanese 
Monbu-Kagaku-Sho Grants (Nos. 15740142 and 16029202).

\end{document}